\documentclass[10pt]{article}

\usepackage{graphicx}
\usepackage{amsmath}
\usepackage{amssymb}
\usepackage{tikz}
\usepackage{url}
\usepackage{authblk}

\usepackage[a4paper, total={6.5in, 8.5in}]{geometry}

\newcommand{\exam}{\begin{flushleft}\scriptsize\begin{tabular}{l}}
\newcommand{\maxe}{\end{tabular}\end{flushleft}}
\newcommand{\plaat}[1]{\begin{figure}[#1]\begin{center}}
\newcommand{\taalp}[2]{\end{center}\caption{#1}\label{#2}\end{figure}}
\newcommand{\tabel}[1]{\begin{table}[#1]\begin{center}}
\newcommand{\lebat}[2]{\end{center}\caption{#1}\label{#2}\end{table}}

\begin{document}

\title{Compiler Support for Sparse Tensor Computations in MLIR}
\author[1]{Aart J.C. Bik}
\author[1]{Penporn Koanantakool}
\author[1]{Tatiana Shpeisman}
\author[1]{Nicolas Vasilache}
\author[1]{Bixia Zheng}
\author[2]{Fredrik Kjolstad}
\affil[1]{Google}
\affil[2]{Stanford University}
\date{Corresponding author: \texttt{ajcbik@google.com}}


\maketitle

\begin{abstract}
Sparse tensors arise in problems in science, engineering, machine learning, and data analytics. Programs that operate on such tensors can exploit sparsity to reduce storage requirements and computational time. Developing and maintaining sparse software by hand, however, is a complex and error-prone task. Therefore, we propose treating sparsity as a property of tensors, not a tedious implementation task, and letting a sparse compiler generate sparse code automatically from a sparsity-agnostic definition of the computation. This paper discusses integrating this idea into MLIR.
\end{abstract}

\section{Introduction}

Vectors, matrices, and their higher dimensional generalization into tensors that contain many zero elements are called sparse tensors. Sparse tensors arise in a wide range of problems in science, engineering, machine learning, and data analytics. Programs that operate on such tensors can exploit sparsity to reduce both storage requirements and computational time by only storing the nonzero elements and skipping computations with a trivial outcome (such as \verb|x+0=x| and \verb|x*0=0|). This exploitation comes at a cost, though, since developing and maintaining sparse software by hand is a rather complex and error-prone task. Therefore, we propose treating sparsity merely as a property of tensors, not a tedious implementation task, and letting a \textbf{sparse compiler} generate sparse code automatically from a sparsity-agnostic definition of the computation.

In this paper, we describe our experience integrating compiler support for sparse tensor computations into the MLIR open-source compiler infrastructure~\cite{mlir2020} and the opportunities and challenges that arise. The idea of making sparsity a property composes well with MLIR's ease of defining multi-level abstractions and progressively lowering transformations that continuously operate at the most appropriate level of abstraction.

The sparse compiler support in MLIR consists of a new sparse dialect that provides the attributes, types, operations, and transformations that are required to make sparse tensor types first class citizens in MLIR. The sparse dialect forms a bridge between high-level operations on sparse tensors types and low-level operations on sparse storage formats that avoid redundant work by only storing and operating on nonzero elements. Central to sparse tensor types is an encoding attribute that defines the desired way of storing sparse tensors. For example, the well-known CSC (Compressed Sparse Column) storage format for sparse matrices can be defined with the following attribute, using a per-dimension \texttt{dense} or \texttt{compressed} level type to indicate full or compressed storage in that dimension and a permutation map to indicate the desired dimension order (here, column-wise).

\exam
\verb"#CSC = #sparse_tensor.encoding<{" \\
\verb+  dimLevelType = [ "dense", "compressed" ],+ \\
\verb"  dimOrdering = affine_map<(i,j) -> (j,i)>" \\
\verb"}>"
\maxe

The dense MLIR representation of a matrix multiplication \texttt{C[i,j] = A[i,k] * B[k,j]}
\exam
\verb"%C = linalg.matmul ins(%A, %B: tensor<?x?xf64>,       tensor<?x?xf64>) -> tensor<?x?xf64>" \\
\maxe
changes into a sparse operation by \emph{merely} adding a sparse attribute to the tensor type of one or more operands, as done below to define a SpMM kernel that uses the CSC storage for matrix \texttt{A}.
\exam
\verb"%C = linalg.matmul ins(%A, %B: tensor<?x?xf64, "\textbf{\texttt{\#CSC}}\verb">, tensor<?x?xf64>) -> tensor<?x?xf64>" \\
\maxe

By making sparsity an optional property of the tensors, we avoid the proliferation of specialized routines for such operations. Instead, after introducing the sparse tensor attribute, compiler transformations take care of lowering the operation to imperative constructs and sparse storage formats that only store and iterate over nonzero elements to perform the matrix multiplication. Modifying the sparse attribute of matrix \texttt{A}, such as preferring row-wise access, or adding sparse attributes to the tensor types of matrices \texttt{B}, \texttt{C}, or both, will prompt the compiler to generate completely different sparse kernels. Since programmers merely annotate sparse tensor types, and leave the tedious implementation task to the sparse compiler, this approach greatly simplifies sparse code development compared to traditional hand-written approaches. Now, a single sparsity-agnostic description can be mapped into a wide range of sparse implementations, each tailored to specific instances of the same problem.

The rest of this paper is organized as follows. Section~\ref{prelim} provides preliminaries related to sparse tensors. Sections~\ref{sparsemlir} and~\ref{sparse_dialect} dive into the design and implementation details of adding sparse compiler support to the MLIR compiler infrastructure. Section~\ref{moda} explores various ways to use the new support in MLIR. An experimental validation of the sparse compiler follows in Section~\ref{experiments}. Related work is discussed in Section~\ref{related}. Finally, conclusions and future plans appear in Section~\ref{conclusions}.

\section{Sparse Preliminaries}
\label{prelim}

This section provides preliminaries on sparse tensors, sparse storage formats, and sparse compilers.

\subsection{Sparse Tensors and Storage Formats}

A \textbf{tensor} is a $d$-dimensional generalization of one-dimensional vectors and two-dimensional matrices. If many elements in the tensor are zero, the tensor is called a \textbf{sparse tensor}, which is a situation that arises often in problems in science, engineering, machine learning, and data analytics. In contrast, a tensor without this property is called a \textbf{dense tensor}. One can furthermore distinguish between \emph{unstructured} sparse tensors that have no discernible nonzero structures and \emph{structured} sparse tensors that have particular nonzero structures, such as tensors with nonzero elements confined within blocks, bands, diagonals, or borders, or with certain statistical properties on the distribution of nonzero elements.

Programs that operate on sparse tensors can take advantage of the sparsity to reduce \emph{storage requirements} by only storing the nonzero elements and \emph{computational time} by skipping trivial operations (such as \verb|x+0=x| and \verb|x*0=0|). How to effectively exploit sparse vectors and matrices has been well-studied in the past for linear algebra problems~\cite{anderson,coleman,curtis71,duffo,duff,duff5,evans,george,harary,mann,sergio,marinus,liu,reid2,tewarson,zlatev}. The growing popularity of deep learning and big data has sparked a similar interest in studying how machine learning kernels can take advantage of sparse tensors~\cite{chou2018,drkjolstad,taco2018,taco2017,senanayake2020,fiber,tew2016}.

Many different ways of storing sparse tensors have been proposed~\cite{sato1963,tinney,buluc2008,itpack,rice1985,saad2003,buluc2009,smith2015a,li2018hicoo} and which of these storage formats is most effective in terms of minimizing storage as well as computation depends on the peculiarities of the nonzero structure, the operations to be performed, and the target architecture. All sparse storage formats consist of a \textbf{primary storage} that stores the numerical values of the nonzero elements and some \textbf{overhead storage} needed to map those values to their tensor coordinates, in order to reconstruct the enveloping tensor from these values. Note that sometimes, even some zero elements are stored explicitly to accommodate for simpler structured formats or to avoid costly removals of nonzero elements that become zero during the computation. Such explicit zeros, however, do not impact the correctness of tensor operations.

Well-known examples of sparse storage formats include coordinate format (COO)~\cite{sato1963}, compressed sparse row/column (CSR and CSC)~\cite{tinney}, doubly compressed sparse row/column (DCSR and DCSC)~\cite{buluc2008}, compressed sparse fiber (CSF)~\cite{smith2015a}, block, band, diagonal and jagged diagonal formats, and hash maps. Although several examples appear later in the paper, for an in-depth survey of sparse storage formats, we must refer to the literature. 

\subsection{Sparse Compilers}

Writing effective sparse code is a time-consuming and error-prone task, further complicated by the large number of possible combinations of storage formats, nonzero structures, operations, and target architectures. Due to the complexity, programmers instead usually restrict themselves to hand-optimizing a small set of library methods for specific operations and storage formats, and building larger sparse programs by composing available library methods. This approach may lead to sub-optimal performance when costly data structure conversions are needed in between library methods or, worse, sub-optimal asymptotic complexity when many avoidable intermediate results produced by one library method are nullified after multiplication by zeros in the next method.

Rather we would like to \emph{treat sparsity merely as a property}, not a tedious implementation task, and let a \textbf{sparse compiler} generate sparse code automatically from a sparsity-agnostic definition of the computation. This idea of keeping sparsity completely transparent to the programmer was pioneered in the \emph{MT1 sparse compiler} for linear algebra~\cite{drbik,biktms,bikjpdc,biktpds} and later formalized and generalized to tensor algebra in \emph{TACO (Tensor Algebra Compiler)}~\cite{taco,drkjolstad,taco2018,taco2017}. With the figurative ``push of a button'', a sparse compiler can convert a single sparsity-agnostic description into a wide range of sparse implementations, each tailored to specific instances of the same problem with tensors stored in different data structures. This automatic approach not only enables non-expert programmers to generate sparse code quickly, but also enables expert programmers to explore the full space of possible sparse implementations.

\section{A Sparse Compiler in MLIR}
\label{sparsemlir}

This section discusses some of our design considerations from implementing a sparse compiler in the MLIR compiler infrastructure, which is part of the LLVM open-source project.

\subsection{MLIR Compiler Infrastructure}

The MLIR open-source project~\cite{mlir2021,mlir2020} provides an extensible infrastructure for building compilers for domain specific languages. The infrastructure provides a way to specify new intermediate representations through \textbf{dialects} together with transformations on these representations. Transformations can be written as compositions of orthogonal localized match and rewrite primitives. These are often decomposed further into \textbf{rewriting rules} when applied within a dialect and \textbf{lowering rules} when converting from a higher-level dialect to a lower-level dialect. Each dialect can define custom attributes, types, and operations. Throughout the compilation, separate dialects can co-exist to form a hybrid representation of a program. The ability to progressively lower to dialects closer and closer to the target hardware during the compilation process, together with an intuitive transformation mechanism, has made MLIR a popular compiler infrastructure for domain specific languages that need to bridge large semantic gaps, such as compiling for machine learning. 

\begin{figure}[ht]
    \centering
    \includegraphics[width=4cm]{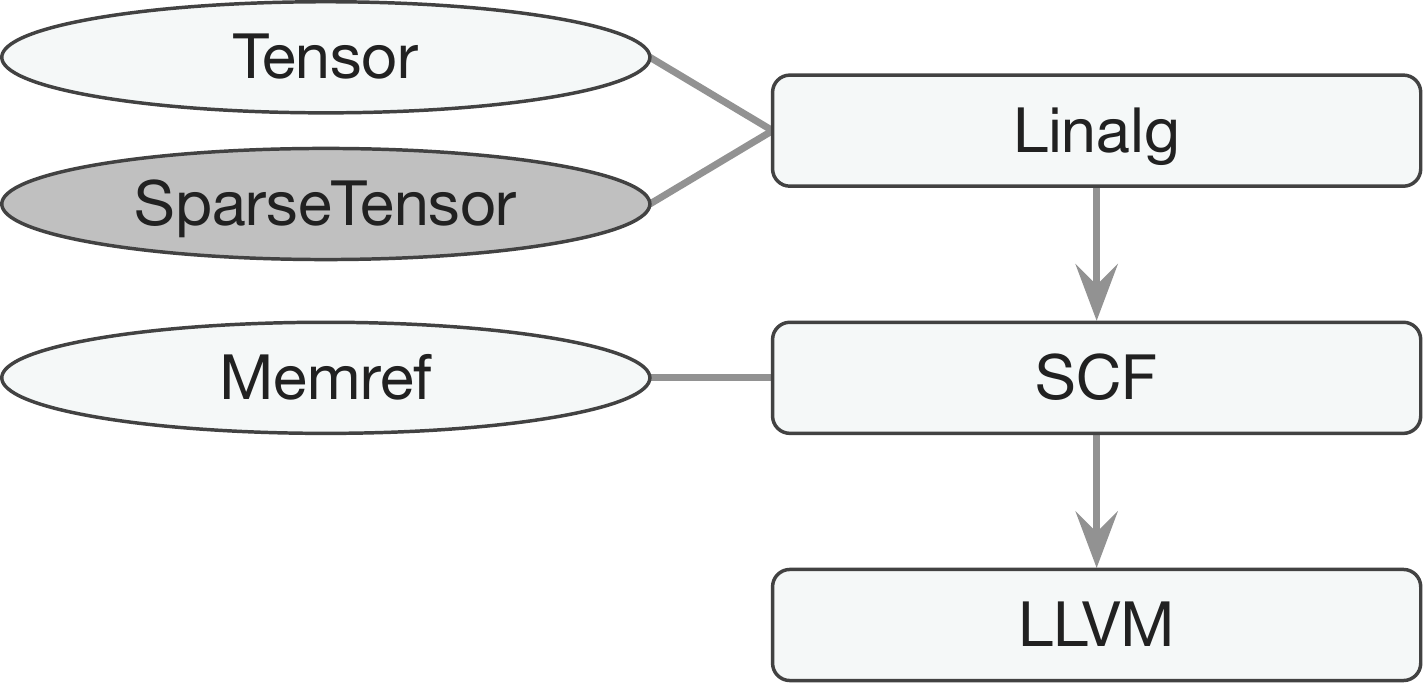}
    \caption{
        The MLIR dialects that are involved in sparse compilation. Ovals contain data, rounded squares contain code, lines indicate data that is used by a code dialect, and arrow indicates a code dialect can be lowered to another code dialect. Although sparse compilation only adds the \texttt{SparseTensor} dialect with a new tensor type, it reuses \texttt{Linalg} to define its operations and \texttt{SCF} as its target.
        \label{fig:overview}
    }
\end{figure}

In this paper, we rely on several MLIR dialects, whose relationships are shown in Figure~\ref{fig:overview}. The dialects are briefly described below, ordered from higher to lower levels of abstraction.

\paragraph{Linalg}
The \texttt{Linalg} dialect, inspired by the tensor index notation found in tensor comprehensions~\cite{nicolas2018}, provides high-level primitives that know how to decompose themselves in a generic fashion and capture structural information useful for maintaining invariants and implementing transformations for high-performance. The central idea is \emph{carrying high-level static information as a first-class citizen in the IR}.
A key design of the current work was to leave the operational semantics of the dialect unchanged, except that it now supports either dense or sparse data types. We assume that a high-level kernel written in, e.g., Tensorflow, JAX, or Python, is presented to the sparse compiler as a sequence of \texttt{Linalg} operations, possibly after lowering from higher-level dialects.

\paragraph{Tensor}
The \texttt{Tensor} dialect operations manipulate an abstract tensor type for which the compiler infrastructure has not yet decided on a representation in memory.  When tensors are small and of static sizes, they may be directly promoted to constants or vectors and can bypass memory to be stored in registers. When tensors are large enough and/or have dynamic sizes, they are assigned memory storage through a \textbf{bufferization} process.

\paragraph{SparseTensor}
The \texttt{SparseTensor} dialect, discussed in more detail in Section~\ref{sparse_dialect}, is new to this work and provides the attributes, types, operations, and transformations that are required to make sparse tensor types first-class citizens within the MLIR compiler infrastructure. The dialect provides a bridge between high-level operations on these sparse tensors types and low-level operations that only store and operate on nonzero elements to avoid performing redundant work. It co-exists with the \texttt{Tensor} dialect to provide a logical separation between purely dense and sparse functionality.

\paragraph{SCF}
The structured control-flow \texttt{SCF} dialect provides operations that represent looping and conditionals, e.g. regular \texttt{for} and \texttt{while} loops (without early exit) as well as an \texttt{if} conditional construct. This is structured at a higher-level of abstraction than control-flow graphs (CFG). Notably, loop operations may yield SSA values and compose well with other operations and dialects with either SSA-based side-effect-free or memory-based side-effecting semantics. This dialect is used by the sparse compiler to decompose sparse \texttt{Linalg} operations into explicit imperative constructs.

\paragraph{Memref}
The \texttt{Memref} dialect introduces the \texttt{memref} data type which is the main representation for memory buffers in MLIR and the entry point to the side-effecting memory-based operations. The \texttt{memref} data type also provides an unsurprising ABI to interoperate with external C code and serves as a  bridge for calling libraries from MLIR codegen. Dense tensors undergo a relatively straightforward bufferization into memory buffers (viz. multi-dimensional arrays). The bufferization of sparse tensors is more elaborate, as will be seen later in this paper. 

\subsection{Sparse Compiler Design Philosophy}

The growing popularity of the MLIR open-source project together with the growing interest in sparse tensors in machine learning and data analytics prompted adding compiler support for sparse tensor computations to the MLIR compiler infrastructure. The idea of making sparsity a property composes well with MLIR's ease of defining multi-level abstractions and progressively lowering transformations that continuously operate at the most appropriate level of abstraction.

The \emph{north star vision} of this project consists of providing an excellent, reusable sparse ecosystem to academia and industry that is based on first principles. To this end, emphasis was put on introducing sparse tensor types as proper first-class citizens into MLIR. Furthermore, a first \emph{reference implementation} provides a fully functional implementation of this sparse ecosystem against which future versions can be compared. The long-term north star vision together with the shorter-term first reference implementation enables researchers to independently start developing MLIR front-ends for array languages that incorporate some form of sparsity annotations as well as new MLIR transformations that implement improved or even alternative sparse compilers for prototyping and production. Our hope is that the open-source nature of MLIR will work both ways, that is, benefit researchers that are new to the sparse domain as well as as solicit useful contributions from experts in the open-source community at large.  

\section{The Sparse Tensor Dialect}
\label{sparse_dialect}

Sparse tensor support in MLIR mostly resides within a new \texttt{SparseTensor} dialect, which provides the attributes, types, operations, and transformations that are required to make sparse tensor types first class citizens within the MLIR compiler infrastructure. The dialect forms a bridge between high-level operations on sparse tensors types and low-level operations on the actual sparse storage formats that only store and operate on nonzero elements to avoid performing redundant work.

\subsection{Sparse Tensor Attributes and Types}

Many choices are possible for the sparse tensor type specification itself, varying from simply annotating a tensor type as sparse to providing detailed information on the nonzero structure or other characteristics. We support a TACO-flavored mechanism of annotating sparse tensors~\cite{drkjolstad}, since this provides an elegant and powerful way of specifying a large number of different sparse storage formats. In true MLIR fashion, however, the type specification has been kept extensible to allow for more advanced annotations in the future.

Central to the chosen annotation mechanism is an encoding attribute that defines the desired way of storing each sparse tensor by means of (1) \emph{per-dimension level types} of either \texttt{dense} or \texttt{compressed}, (2) a \emph{dimension ordering}, and (3) \emph{bit widths} for pointers and indices.
\exam
\verb"#SparseTensor = #sparse_tensor.encoding<{" \\
\verb+  dimLevelType = [ "dense", "compressed", ... ],+ \\
\verb"  dimOrdering = affine_map<(i,j,k,...) -> (i,j,k,...)>," \\
\verb"  pointerBitWidth = ...," \\
\verb"  indexBitWidth = ...," \\
\verb"}>"
\maxe

Sparse tensor types are obtained by annotating the built-in tensor types of MLIR with sparse format attributes as illustrated below.

\exam
\verb"tensor<..., #SparseTensor>"
\maxe

The attribute indicates that when the $d$-dimensional tensor is lowered into an actual storage scheme during bufferization, rather than selecting a straightforward \texttt{memref} buffer used for dense tensors, the sparse tensor is lowered into a compact sparse storage scheme that \emph{conceptually} consists of two integral jagged arrays \texttt{pointers[}$d$\texttt{][*]} and \texttt{indices[}$d$\texttt{][*]}, where each entry in the first dimension stores, respectively, the pointer and index array of a single tensor dimension, and a one-dimensional values array \texttt{values[*]} that provides sufficient space for all stored elements.

Consider, for example, the following sparse vector type.
\exam
\verb"#SparseVector= #sparse_tensor.encoding<{" \\
\verb+   dimLevelType = [ "compressed" ]+ \\
\verb"}>" \\
\verb"tensor<16xf64, #SparseVector>"
\maxe
Then, the sparse vector\footnote{
Tensors in this paper use zero-based indexing, e.g. a three-dimensional tensor $T$ starts with element $t_{0,0,0}$ with, from left-to-right, layer, row, and column index~0.
}
\[ \footnotesize
\vec{x} = 
 (  0, 0, 0, x_{3}, 0, 0, x_{6}, x_{7}, 0, 0, x_{10}, 0, 0, 0, 0, 0 )
\]
has the following compressed format results. Contents of \texttt{pointers[0][0]} and \texttt{pointers[0][1]} denote that elements appear at range \texttt{[0,4)} in the indices and values arrays. There, \texttt{indices[0][*]} and \texttt{values[*]} provide, respectively, the indices and values of nonzero elements. Memory is saved by only storing 10 values (4 doubles primary and 6 integers overhead storage) rather than the 16 consecutive doubles that would result in the dense case.

\exam
\begin{tabular}{l|c|c|c|c|}
\cline{2-3}
\texttt{pointers[0]:} & 0 & 4 \\
\cline{2-5}
\texttt{indices[0]:}  & 3 & 6 & 7 & 10 \\
\cline{2-5}
\texttt{values:}      & $x_{3}$ & $x_{6}$ & $x_{7}$ & $x_{10}$ \\
\cline{2-5}
\end{tabular}
\maxe

For matrices, compressed sparse row (CSR) is defined with dimension level type \texttt{dense} for the first dimension and \texttt{compressed} for the second dimension, as follows. When absent, the dimension ordering defaults to lexicographic index order, thus, row-wise order in this case.

\exam
\verb"#CSR = #sparse_tensor.encoding<{" \\
\verb+  dimLevelType = [ "dense", "compressed" ]+ \\
\verb"}>" \\
\verb"tensor<3x4xf64, #CSR>" \\
\maxe
Then, the following sparse matrix

\[ \footnotesize
A = \begin{pmatrix}
a_{0,0} & 0 & 0 & a_{0,3} \\
0       & 0 & 0 & 0 \\
a_{2,0} & 0 & 0 & 0 \\
\end{pmatrix}
\]
maps to the following compressed format which has an implicit dense first dimension of size~3 and only uses the second dimension of the jagged arrays. The nonzero values and corresponding column coordinates in each row~$0 \leq \texttt{i} < 3$ can be found in the indices and values arrays at range \texttt{pointers[1][i]} up to \texttt{pointers[1][i+1]}. There, \texttt{indices[1][*]} and \texttt{values[*]} provide, respectively, column indices and values. Note that the number 2 is repeated in the pointers array because the second row of the matrix $A$ is empty.

\exam
\begin{tabular}{l|c|c|c|c|}
\cline{2-5}
\texttt{pointers[1]:} & 0 & 2 & 2 & \ 3 \ \\
\cline{2-5}
\texttt{indices[1]:}  & 0 & 3 & 0 \\
\cline{2-4}
\texttt{values:}      & $a_{0,0}$ & $a_{0,3}$ & $a_{2,0}$ \\
\cline{2-4}
\end{tabular}
\maxe

Alternatively, using \texttt{compressed} for the first dimension and \texttt{dense} for the second dimension maps to the following storage format which only uses the first dimension of the jagged arrays and contains some explicit zeros in the, now, fully stored dense rows with implicit size~4 for the second dimension. This storage scheme only skips one row in this case, but may favor target architectures with efficient vector instructions.
\exam
\begin{tabular}{l|c|c|c|c|c|c|c|c|}
\cline{2-3}
\texttt{pointers[0]:} & 0 & 2 \\
\cline{2-3}
\texttt{indices[0]:}  & 0 & 2 \\
\cline{2-9}
\texttt{values:}      & $a_{0,0}$ & \ 0 \ & \ 0 \ & $a_{0,3}$ &
                        $a_{2,0}$ & \ 0 \ & \ 0 \ & \ 0 \ \\
\cline{2-9}
\end{tabular}
\maxe

The storage format can be made doubly compressed using \texttt{compressed} for both dimension level types. In addition, an explicit \emph{dimension ordering} can be used to enforce column-wise storage over the default lexicographic index order storage, as illustrated below with the definition of doubly compressed sparse column (DCSC).

\exam
\verb"#DCSC = #sparse_tensor.encoding<{" \\
\verb+  dimLevelType = [ "compressed", "compressed" ],+ \\
\verb"  dimOrdering  = affine_map<(i,j) -> (j,i)>" \\
\verb"}>" \\
\verb"tensor<3x4xf64, #DCSC>" \\
\maxe
This type maps to the following column-wise sparse storage format for the example matrix $A$. The contents in \texttt{pointers[0][0]} and \texttt{pointers[0][1]} denote that range \texttt{[0,2)} is used in the next arrays to store two columns. There, \texttt{indices[0][*]} denote that only column~0 and column~3 contain nonzero elements. The next level provides a similar structure within each explicitly stored column.

\exam
\begin{tabular}{l|c|c|c|c|}
\cline{2-3}
\texttt{pointers[0]:}  & 0 & 2 \\
\cline{2-3}
\texttt{indices[0]:}  & 0 & 3 \\
\cline{2-4}
\texttt{pointers[1]:} & 0 & 2 & 3 \\
\cline{2-4}
\texttt{indices[1]:}  & 0 & 2 & 0 \\
\cline{2-4}
\texttt{values:}    & $a_{0,0}$ & $a_{2,0}$ & $a_{0,3}$ \\
\cline{2-4}
\end{tabular}
\maxe

The encoding attribute easily generalizes to tensor or arbitrary dimensions, as illustrated below for a 3-dimensional tensor.

\exam
\verb"#SparseTensor = #sparse_tensor.encoding<{" \\
\verb+  dimLevelType = [ "compressed", "compressed", "compressed" ]+ \\
\verb"}>" \\
\verb"tensor<3x3x4xf64, #SparseTensor>" \\
\maxe
Consider the tensor $T$ shown below, with layers ordered front to back.

\begin{flushleft}
\footnotesize
\begin{tikzpicture}
 \node at (0,0)(mm1){$
  \begin{pmatrix}
   t_{0,0,0} & \ 0 \ & \ 0 \ & \ 0 \ \\
   \ 0 \ & \ 0 \ & \ 0 \ & \ 0 \ \\
   \ 0 \ & \ 0 \ & \ 0 \ & \ 0 \ \\
  \end{pmatrix}
 $};
 \node at (1.1,1.1)(mm2){$
 \begin{pmatrix}
   \ 0 \ & \ 0 \ & \ 0 \ & \ 0 \ \\
   \ 0 \ & \ 0 \ & \ 0 \ & \ 0 \ \\
   \ 0 \ & \ 0 \ & \ 0 \ & \ 0 \ \\
 \end{pmatrix}
 $};
\node at (2.2,2.2)(mm3){$
 \begin{pmatrix}
   t_{2,0,0} & \ 0 \  & t_{2,0,2} & \ 0 \ \\
   \ 0 \ & \ 0 \ & t_{2,1,2} & t_{2,1,3} \\
   \ 0 \ & \ 0 \ & \ 0 \ & \ 0 \ \\
 \end{pmatrix}
 $};
\draw [dotted] (-1.3,  0.7) -- (0.2, 2.7);
\draw [dotted] ( 1.8, -0.5) -- (3.3, 1.5);
\end{tikzpicture}
\end{flushleft}

With the triply-compressed encoding attribute shown above, the following compressed data structure results. Stored elements appear in the default lexicographic index order. Contents in \texttt{pointers[0][0]} and \texttt{pointers[0][1]} reveal two filled layers, with \texttt{indices[0][*]} specifying layer~0 and~2 in particular. The next level \texttt{pointers[1][*]} shows that the matrix in the first stored layer has one filled row and the matrix in the second stored layer has two filled rows. The last level provides structure within each compressed row.

\exam
\begin{tabular}{l|c|c|c|c|c|}
\cline{2-3}
\texttt{pointers[0]:}  & 0 & 2 \\
\cline{2-3}
\texttt{indices[0]:}  & 0 & 2 \\
\cline{2-4}
\texttt{pointers[1]:} & 0 & 1 & 3 \\
\cline{2-4}
\texttt{indices[1]:}  & 0 & 0 & 1 \\
\cline{2-5}
\texttt{pointers[2]:} & 0 & 1 & 3 & 5 \\
\cline{2-6}
\texttt{indices[2]:}  & 0 & 0 & 2 & 2 & 3 \\
\cline{2-6}
\texttt{values:}    & $t_{0,0,0}$ & $t_{2,0,0}$ & $t_{2,0,2}$ & $t_{2,1,2}$ & $t_{2,1,3}$ \\
\cline{2-6}
\end{tabular}
\maxe

In addition to the per-dimension level types and optional dimension ordering, the encoding attribute can also specify alternative bit-widths ranging from 8 to 64 for pointers and indices, defaulting to 0 to denote the native bit-width of the target architecture. Narrower bit-widths can be used to reduce overhead storage requirements even further, provided that both widths still suffice to store the maximum possible integral pointer and index value throughout the full storage format.
Given the two choices for each per-dimension level type, all possible permutations for the dimension ordering, and the four choices for both bit-widths, this relatively simply sparse tensor type encoding already supports $2^d \cdot d! \cdot 16$ different ways of storing a single $d$-dimensional tensor. Nevertheless, as stated before, the encoding has been kept extensible to allow for specifying even more sparse storage formats in the future, such as the new dimension level types discussed in~\cite{chou2018}.

\subsection{Sparse Tensor Operations}

With sparse tensor types as proper first-class citizens, the idea is that \emph{any} MLIR tensor operation can be made sparse by simply annotating the tensor types of the operands. For example, the \texttt{Linalg} dialect is an intermediate representation commonly used by MLIR to progressively lower machine learning kernels to actual executable code. The dense MLIR representation of a matrix multiplication operation \texttt{C[i,j] = A[i,k] * B[k,j]}

\exam
\verb"%C = linalg.matmul ins(%A, %B: tensor<?x?xf64>, tensor<?x?xf64>) -> tensor<?x?xf64>" \\
\maxe
changes into a sparse kernel by \emph{merely} adding sparse attributes to the tensor types of the operands, as was already shown in the introduction for SpMM. Here, we convert the operation into a SpMSpM kernel as follows, where all matrices are stored in CSR format.

\exam
\verb"%C = linalg.matmul ins(%A, %B: tensor<?x?xf64, #CSR>, tensor<?x?xf64, #CSR>) -> tensor<?x?xf64, #CSR>" \\
\maxe

As such, the \texttt{SparseTensor} dialect only needs to provide a few operations specific to sparse tensor types: (1) materialization operations, (2) conversion operations, (3) operations that support progressive lowering. Below, we briefly explore each category.

The \texttt{new} operation materializes a sparse tensor from some source. 

\exam
\verb"%0 = sparse_tensor.new %source : !Source to tensor<?x?x?xf32, #SparseTensor>"
\maxe

Most commonly, the source is a file in one of the external formats provided by the Matrix Market~\cite{matrixmarket} (a popular sparse matrix repository and successor of earlier sets like the Harwell-Boeing Sparse Matrix Collection~\cite{duff2} and SPARSKIT~\cite{saad2}), or FROSTT~\cite{frostt} (a repository of open sparse tensors). Reading from file assumes a target platform that supports a file system, but other more architecturally neutral sources are possible as well, such as initializing a sparse tensor from code or through some API from a memory-resident format owned by an external library. Another materialization operation is the \texttt{init} operation, which materializes an uninitialized tensor into a computation.
Conversely, the \texttt{out} operation, illustrated below, dematerializes a sparse tensor to some destination, typically by writing an external format like FROSTT to a file if the target architecture provides a file system. The operation is kept general, however, to support other kinds of destinations in the future as well.

\exam
\verb"sparse_tensor.out %0, %dest : tensor<?x?x?xf32, #SparseTensor>, !Destination" \\
\maxe

Since MLIR is strongly typed, built-in verification rejects implicit type casts between dense and sparse tensor types or between differently annotated sparse tensor types, since most of such type conversions would incur non-trivial costs. Instead, conversions that involve sparse tensor types must be made explicit with the \texttt{convert} operation.

An example of converting a $10 \times 10$ sparse matrix in CSR format to CSC format with dynamic dimensions is shown below.

\exam
\verb"%to = sparse_tensor.convert %from : tensor<10x10xf64, #CSR> to tensor<?x?xf64, #CSC>"
\maxe

In the first reference implementation, conversion operations between sparse formats are implemented by converting to and from an intermediate coordinate scheme, which avoids the quadratic trap of implementing all possible direct conversions. In the longer term, however, this approach can be replaced by more advanced schemes, at least in special cases, such as proposed in~\cite{chou2020}.

The \texttt{convert} operation also converts between dense tensor types and sparse tensor types. Although not practical for many real-world sparse tensors, where storage requirements of an enveloping dense tensor would be prohibitive, such conversions are useful to manipulate smaller data sets in tests. An example of initializing a sparse vector with only the nonzero integers from a compile-time \texttt{dense} constant vector is shown below.

\exam
\verb"%v = arith.constant dense<[ 0, 0, 2, -1, 0, 0, 2, 0, 0, 0 ]> : tensor<10xi32>" \\
\verb"%s = sparse_tensor.convert %v : tensor<10xi32> to tensor<10xi32, #SparseVector>" \\
\maxe

MLIR also provides a compile-time \texttt{sparse} constant that uses a COO-flavored mechanism of specifying the indices and values of all nonzero elements in two different lists, as shown below.
Although, perhaps somewhat surprising, the type of SSA value \texttt{\%m} is actually a \emph{dense} $10 \times 8$ matrix, the subsequent conversion ensures that the \emph{sparse} matrix in \texttt{\%s} is initialized directly from the nonzero elements, without ever fully materializing the enveloping dense matrix at runtime.

\exam
\verb"%m = arith.constant sparse<" \\
\verb"  [ [0, 0], [0, 7], [1, 2], [4, 2], [5, 3], [6, 4], [6, 6], [9, 7]]," \\
\verb"    [1.0, 2.0, 3.0, 4.0, 5.0, 6.0, 7.0, 8.0]> : tensor<10x8xf64>" \\
\verb"%s = sparse_tensor.convert %m : tensor<10x8xf64> to tensor<10x8xf64, #CSR>" 
\maxe

Lastly, the \texttt{SparseTensor} dialect provides operations that facilitate progressively lowering, first from annotated kernels to an intermediate form that only stores and iterates over nonzero elements, without fully committing to an underlying storage format yet, then from this intermediate form to a bufferized form with actual sparse storage schemes, and then finally to executable code that runs on the target hardware. In the intermediate form, it is useful to have some primitives that provide sparsity-specific functionality while still hiding full implementation details of the underlying sparse storage formats.
For instance, the \texttt{indices} operation returns the indices array of the sparse storage format at the given dimension for the given sparse tensor through a one-dimensional array, called a \texttt{memref} after bufferization. An example is shown below.

\exam
\verb"%ind = sparse_tensor.indices %t, %c3 : tensor<10x10x10x10xf64, #SparseTensor> to memref<?xindex>" \\
\maxe

Other operations related to sparsity consist of obtaining the pointers or values array, and  inserting elements into a sparse tensor, possibly assisted by expanding and compressing access patterns within innermost loops, as further explained in the next section.

\subsection{Sparse Tensor Transformations}
\label{trafo}

The sparse compiler itself is implemented as a set of lowering transformations that reside in the \texttt{SparseTensor} dialect. These transformations take care of lowering annotated kernels, i.e. operations on sparse tensor types, to imperative constructs and sparse storage formats that only store and iterate over the nonzero elements, possibly optimized with parallelization or vectorization of the generated loops. In true MLIR fashion, transformation definitions separate \textbf{mechanism}, i.e. how to modify an intermediate form, from \textbf{policy}, i.e. which transformations should be applied and in what order to get the best performing result.

The MLIR sparse compiler transformations closely follow the \emph{sparse iteration model} of TACO~\cite{drkjolstad}. Starting with a tensor index notation expressed in the \texttt{Linalg} dialect, a topologically sorted iteration graph is constructed that reflects the required order on the implicit indices with respect to the dimension ordering of each sparse tensor. Next, iteration lattices are constructed for the tensor expression for every index in topological order. Each iteration lattice point consists of a conjunction of tensor indices together with a tensor (sub)expression that drive the actual sparse code generation, consisting of a straightforward mapping to loops and conditionals using the \texttt{SCF} dialect. The resulting sparse intermediate form that only (co)iterates over the nonzero elements is subsequently bufferized to another intermediate form with actual sparse storage schemes, and then finally lowered to executable code that runs on the target hardware, typically by handing off LLVM IR to the LLVM back-end compiler.

As a simple example, consider the following straightforward \texttt{Linalg} operation that scales the elements in a vector by a constant in-place, viz. \texttt{x[i] *= c}.

\exam
\verb"%0 = linalg.generic #trait_scale" \\
\verb"  outs(%vecx: tensor<?xf64, #SparseVector>) {" \\
\verb"  ^bb(%x: f64):" \\
\verb"    %1 = arith.mulf %x, %c : f64" \\
\verb"    linalg.yield %1 : f64" \\
\verb"  } -> tensor<?xf64, #SparseVector>" \\
\maxe

The steps discussed above reveal that the loop-body only needs to execute for nonzero elements in the sparse vector. This yields the following intermediate form that updates these elements in-place using primitives of the \texttt{SparseTensor} and \texttt{SCF} dialects. 

\exam
\verb"%x_pointers = sparse_tensor.pointers %vecx, %c0" \\
\verb"%x_values = sparse_tensor.values %vecx" \\
\verb"%x_lo = memref.load %x_pointers[%c0] : memref<?xindex>" \\
\verb"%x_hi = memref.load %x_pointers[%c1] : memref<?xindex>" \\
\verb"scf.for %ii = %x_lo to %x_hi step %c1 {" \\
\verb"  %0 = memref.load %x_values[%ii] : memref<?xf64>" \\
\verb"  %1 = arith.mulf %0, %c : f64" \\
\verb"  memref.store %1, %x_values[%ii] : memref<?xf64>" \\
\verb"}"
\maxe

The sparse iteration model also provides an elegant way to generate code that implements \emph{co-iteration}, i.e. iterating over a disjunction or a conjunction of several sparse data structures simultaneously. Consider, for example, computing the inner product \texttt{x += a[i] * b[i]} of two sparse vectors, represented by the following \texttt{Linalg} operation.

\exam
\verb"%0 = linalg.generic #trait_inner" \\
\verb"  ins(%veca, %vecb: tensor<?xf64, #SparseVec>, tensor<?xf64, #SparseVec>) outs(%acc: tensor<f64>) {" \\
\verb"    ^bb(%a: f64, %b: f64, %x: f64):" \\
\verb"      %0 = arith.mulf %a, %b : f64" \\
\verb"      %1 = arith.addf %x, %0 : f64" \\
\verb"      linalg.yield %1 : f64" \\
\verb"} -> tensor<f64>" \\
\maxe

Then following the sparse iteration model reveals that the loop-body only needs to execute when both elements of the sparse vectors are nonzero. This eventually yields in the following \texttt{SCF} construct, presented in a reduced and simplified form to enhance readability (mostly by omitting some of the machinery needed to correctly represent reduction and induction cycles in SSA).

\exam
\verb"%a_lo = memref.load %a_pointers[%c0] : memref<?xindex>" \\
\verb"%a_hi = memref.load %a_pointers[%c1] : memref<?xindex>" \\
\verb"%b_lo = memref.load %b_pointers[%c0] : memref<?xindex>" \\
\verb"%b_hi = memref.load %b_pointers[%c1] : memref<?xindex>" \\
\verb"... = scf.while (...) : (index, index, f64) -> (index, index, f64) {" \\
\verb"  %cond = %ia < %a_hi && %ib < %a_hi" \\
\verb"  scf.condition(%cond)" \\
\verb"} do {" \\
\verb"  ..." \\
\verb"  %ixa = memref.load %a_indices[%ia] : memref<?xindex>" \\
\verb"  %ixb = memref.load %b_indices[%ib] : memref<?xindex>" \\
\verb"  %i = min(%ixa, %ixb)" \\
\verb"  %acc = scf.if (%ixa == %i && %ixb == %i) -> (f64) {" \\
\verb"    %0 = memref.load %a_values[%ia] : memref<?xf64>" \\
\verb"    %1 = memref.load %b_values[%ib] : memref<?xf64>" \\
\verb"    %2 = arith.mulf %0, %1 : f64" \\
\verb"    %3 = arith.addf %acc_in, %2 : f64" \\
\verb"    scf.yield %3 : f64" \\
\verb"  } else {" \\
\verb"    scf.yield %acc_in : f64" \\
\verb"  }" \\
\verb"  update %ia++ if %ixa == %i" \\
\verb"  update %ib++ if %ixb == %i" \\
\verb"}" \\ 
\maxe

Sparse tensor outputs are handled with direct insertions in pure lexicographical index order if all loops that correspond to left-hand-side indices are outermost. Otherwise, the sparse compiler uses a technique called access pattern expansion or workspace~\cite{drbik,duff,rose,taco2018,sergio,pugh1998sipr}, which is a well-known way of efficiently dealing with sparse insertions as well as an alternative way to implement co-iteration over a conjunction or disjunction.
Consider, for instance, the following SpMSpM kernel for \texttt{C[i,j] = A[i,k] * B[k,j]}.

\exam
\verb"%C = linalg.matmul ins(%A, %B: tensor<?x?xf64, #CSR>, tensor<?x?xf64, #CSR>) -> tensor<?x?xf64, #CSR>"
\maxe

With row-wise storage for all matrices, the reduction loop cannot appear innermost. As a result, the sparse compiler transformations lower this kernel to the following sparse intermediate form, again presented in a slightly reduced and simplified form to enhance readability.

\exam
\verb"scf.for %i = %c0 to %m step %c1 {" \\
\verb"  %c_i_values, %filled, %added, %count = sparse_tensor.expand %C" \\
\verb"  ..." \\
\verb"  scf.for %kk = %a_lo to %a_hi step %c1 {" \\
\verb"    %k = memref.load %a_indices[%kk] : memref<?xindex>" \\
\verb"    %aik = memref.load %a_values[%kk] : memref<?xf64>" \\
\verb"    ..." \\
\verb"    scf.for %jj = %b_lo to %b_hi step %c1 {" \\
\verb"      %j = memref.load %b_indices[%jj] : memref<?xindex>" \\
\verb"      %bkj = memref.load %b_values[%jj] : memref<?xf64>" \\
\verb"      %cij = memref.load %c_i_values[%j] : memref<?xf64>" \\
\verb"      %0 = arith.mulf %aik, %bkj : f64" \\
\verb"      %1 = arith.addf %cij, %0 : f64" \\
\verb"      scf.if (not %filled[%j]) {" \\
\verb"        memref.store %true, %filled[%j] : memref<?xi1>" \\
\verb"        record insertion as %added[ %count++ ] = %j" \\
\verb"      }" \\
\verb"      memref.store %1, %c_i_values[%j] : memref<?xf64>" \\
\verb"    }" \\
\verb"  }" \\
\verb"  sparse_tensor.compress %C, %c_i_values, %filled, %added, %count" \\ 
\verb"}" \\
\maxe

The \texttt{expand} operation yields two arrays \texttt{c\_i\_values} and \texttt{filled} with sizes that suffice for a dense innermost dimension (viz. a full row of \texttt{C}). Array \texttt{added} and scalar \texttt{count} are used to keep track of new indices when a false value is encountered in the \texttt{filled} array. The internal allocation and initialization should be done before the loop (preferably even shared between loops of different kernels) so that these dense assignments are amortized over many iterations. Resetting the dense arrays in the loop nest itself is kept sparse by only iterating over set elements through an indirection during the \texttt{compress} operation, so that that amount of work is kept proportional to the number of nonzero elements. This operation also sorts the indices of the new entries, which enables insertions in pure lexicographical index order without costly data movement. 

\section{Sparse Compiler Usage}
\label{moda}

The previous section gave details on the design and implementation of sparse compiler support in the MLIR compiler infrastructure. This section explores two ways to use this new support: end-to-end support for array languages and sparse state space search for testing or library development.

\subsection{End-to-End Sparse Compiler Support for Array Languages}

\begin{figure}[bh]
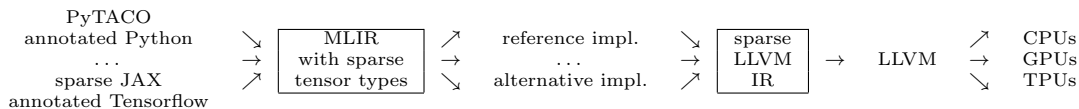

\centering
\scriptsize
\begin{tabular}{cc|c|ccc|c|cccc}
PyTACO \\
\cline{3-3}\cline{7-7}
annotated Python & $\searrow$    & MLIR         & $\nearrow$    & reference impl.   & $\searrow$    & sparse &               &      &  $\nearrow$    & CPUs \\
$\dots$          & $\rightarrow$ & with sparse  & $\rightarrow$ & $\dots$           & $\rightarrow$ & LLVM   & $\rightarrow$ & LLVM &  $\rightarrow$ & GPUs \\
sparse JAX       & $\nearrow$    & tensor types & $\searrow$    & alternative impl. & $\nearrow$    & IR     &               &      &  $\searrow$    & TPUs \\
\cline{3-3}\cline{7-7}
annotated Tensorflow \\
\end{tabular}
\caption{
Overview of retargetable sparse compiler support for array languages. Multiple front-ends map different array languages with sparse annotations, such as PyTACO, to a shared high-level intermediate representation, such as \texttt{Linalg} with sparse tensor types. Alternative sparse compiler pipelines, such as the reference implementation, lower this intermediate to imperative constructs and sparse storage formats that only store and iterate over nonzero elements. This intermediate is handed off to a retargetable backend compiler, like LLVM, that can generate code for a variety of target architectures.
}
\label{fig:end2end}
\end{figure}

Our long-term north star vision centered around sparse tensors types as first class citizens enables researchers to independently start developing front-ends for array languages that incorporate some form of sparsity annotations as well as adding new MLIR transformations that implement improved sparse compilers. This vision would result in the retargetable approach illustrated in Figure~\ref{fig:end2end}, where several front-ends and several sparse compilers share the MLIR infrastructure. All paths eventually generate sparse code in some intermediate form, such as LLVM IR, which can be handed off to a back-end compiler, like LLVM, to generate code for various target architectures.

The end product enables non-expert programmers to exploit sparsity easily, since, by merely adding annotations, a single sparsity-agnostic program in, e.g., Python, could map to a wide range of sparse implementations, each tailored to specific instances of the same problem, both in terms of sparsity properties as well as characteristics of the target architecture. MLIR ships with an initial reference implementation of the sparse compiler transformations. In addition, to illustrate end-to-end support for an array language, we added support for PyTACO~\cite{taco}, which is TACO's Python-based domain specific language for expressing sparse tensor algebra computations.

For example, the following PyTACO code expresses an element-wise matrix addition. The first three lines define tensors \texttt{A} and \texttt{B}, and \texttt{C}. The fourth line defines the two index variables \texttt{i} and \texttt{j}. The last line describes the actual computation, using a tensor index notation to describe how each element in the result tensor can be computed from elements in the two source tensors.

\exam
\verb"A = tensor([1024, 1024], [compressed, dense])" \\
\verb"B = tensor([1024, 1024], [compressed, dense])" \\
\verb"C = tensor([1024, 1024], [compressed, dense])" \\
\verb"i, j = get_index_var(2)" \\
\verb"C[i,j] = A[i,j] + B[i,j]" \\
\maxe

As another example, a matrix multiplication can be expressed as follows.

\exam
\verb"C[i,j] = A[i,k] * B[k,j]" \\
\maxe

PyTACO supports implicit broadcast and implicit reduction. If an index variable is defined in the iteration space but does not appear in the expression of a tensor operand, the tensor operand is broadcast along the dimension represented by the index variable. If an index variable appears in the expression of some tensor operands but not in the expression of the destination tensor, then the corresponding dimension is reduced on the smallest sub-expression that captures the use of the index variable.

The following code illustrates these two rules.

\exam
\verb"C[i,j] = A[i,j] + B[i]         =>  C[i,j] = A[i,j] + broadcast(j, B[i])" \\ 
\verb"D[i] = A[i,j] + B[i,j] + C[i]  =>  D[i] = sum(j, A[i,j] + B[i,j]) + C[i]" \\
\maxe

To support PyTACO, we implemented Python classes for sparsity annotations, index variables, tensors, tensor accesses and tensor expressions. Then, we implemented the dunder method \texttt{\_\_getitem\_\_} to process tensor accesses. A tensor access is a leaf node and a trivial tensor expression. Tensor expressions can participate in operations, such as additions and multiplications, to construct more complicated expressions. We also implemented the dunder method \texttt{\_\_setitem\_\_} to assign a left-hand-side tensor expression to the right-hand-side tensor. 

When a tensor with an assignment is evaluated, we generate the MLIR representation for the tensor assignment. The \texttt{Linalg} dialect has a generic operation that can be used to express the PyTACO tensor computation with only one difference. That is, a reduction in the generic operation is performed on the whole expression, not on the smallest expression that captures the use of the index variables as in the PyTACO. As such, we need to break the following PyTACO expression
\exam
\verb"D[i] = A[i,j] + B[i,j] + C[i]" \\
\maxe
into two generic operations
\exam
\verb"T[i] = A[i,j] + B[i,j]" \\
\verb"D[i] = T[i] + C[i]"
\maxe
to implement the PyTACO expression correctly. When we need to introduce such temporary tensors, we use heuristics to estimate the sparsity for each dimension in the temporary tensor. In particular, if an operation only preserves zero values from both source operands, such as an addition, we choose \texttt{compressed} format for a destination dimension only if both source dimensions are \texttt{compressed}. If an operation preservers zero values from either source operand, such as a multiplication, we choose \texttt{compressed} format for a destination dimension if at least one of its source dimensions is \texttt{compressed}.
After generating the MLIR code for the tensor assignment, we invoke MLIR compilation passes, including the sparse tensor code generator, to compile the code down to a runnable representation. We then invoke the MLIR JIT execution engine to execute this runnable and retrieve the result for the tensor.

\subsection{Sparse State Space Search}

The MLIR infrastructure provides a Python interface for building, compiling, and running MLIR IR from code. This interface provides a convenient way to loop over all possible sparse storage formats and compiler optimizations for a particular kernel. Given the SpMM kernel in which only one matrix is sparse, for example, the code below can be used to exhaustively explore all possible sparsity attributes \texttt{attr} and compiler strategies \texttt{opt} for building and running the kernel.

\exam
\verb"for level in [ [dense, dense], [dense, compressed], [compressed, dense], [compressed, compressed] ]:" \\
\verb"  for ordering in [ ir.AffineMap.get_permutation([0, 1]), ir.AffineMap.get_permutation([1, 0]) ]:" \\
\verb"    for ptrWidth in [0, 8, 16, 32, 64]:" \\
\verb"      for indxWidth in [0, 8, 16, 32, 64]:" \\
\verb"        attr = st.EncodingAttr.get(level, ordering, ptrWidth, idxWidth)" \\
\verb"        mlir = buildSpMM(attr)" \\
\verb"        for opt in compilerStrategies:" \\
\verb"          exec = compileKernel(mlir, opt)" \\
\verb"          result = runKernel(exec, input)" \\
\verb"          verify(result)" \\
\maxe

This approach is useful to stress test the actual sparse compiler implementation, since the computed result should be independent of the actual sparsity annotations or compiler optimizations used (at least within acceptable numerical differences due to floating-point reassociation). However, the same approach can also be used by an expert programmer to exhaustively explore the performance of a library method within the full state space of possible sparse implementations and optimizations before deciding on which one to ship into production.

The loop nest above shows that choosing a proper format for a \emph{single} sparse matrix already gives rise to a state space of 200 configurations. For kernels with higher dimensional or multiple sparse tensors, the size of the state space grows further. Then, when combined with different compiler optimization strategies, different target architectures, and different characteristics of the sparse tensor inputs, the combinatorial explosion of the state space becomes apparent. Although long off-line running times are acceptable while exhaustively searching for the best possible implementation of a library method that will be widely used, at some point exploring the full state space may become infeasible, and the programmer may have to resort to using machine learning for this exploration~\cite{gus}.

\section{Experimental Results}
\label{experiments}

Even though our long term objective is to provide an excellent, reusable sparse ecosystem that simplifies sparse code generation for a wide variety of target architectures (CPUs, GPUs, and TPUs), MLIR ships with a first reference implementation that provides a fully functional implementation of the sparse ecosystem (for CPUs). In this section, we presents experimental validation of this initial reference implementation compared to the state-of-the-art TACO compiler~\cite{taco}. All measurements were taken on an Intel Xeon W2135 3.7GHz.

\subsection{Sparse Tensor Input}

To validate the quality of the reference implementation for reading tensors from files, we measured the time taken to read and pack tensors to an all-dimensions \texttt{compressed} format for a few tensors with varying dimensions from the Matrix Market~\cite{matrixmarket} and FROSTT~\cite{frostt}. For TACO, we used the built-in timing benchmark which invokes GCC during execution to compile generated code. For fair comparison, we compensated the reports to exclude compilation time. For MLIR, we used identical timing methods but the LLVM JIT compiler as back-end. Also, since MLIR reads tensors from an extended FROSTT format with extra size metadata in the header, it has a slight advantage pre-allocating data structures while reading. Table~\ref{tab:sparseio} shows the results, with the relative performance improvement of MLIR over TACO for the total reading and packing time in the last column. The latter gives clear evidence that the reference implementation provides state-of-the-art performance for reading sparse tensors.

\tabel{ht}
\footnotesize
\begin{tabular}{|lrr|rr|rr|l|}
\hline
\multicolumn{3}{|c|}{}              &
\multicolumn{2}{|c|}{\textbf{MLIR}} &
\multicolumn{2}{|c|}{\textbf{TACO}} & \\
\textbf{tensor} & \textbf{dim}  & \textbf{nnz} & 
\textbf{read}   & \textbf{pack} &
\textbf{read}   & \textbf{pack} & \\
\hline
fidap011  & 2 &  1,091,362 &  228 &  136 &   262 &  142 & 1.1x \\
nell-2    & 3 & 76,879,419 & 9568 & 2055 & 13303 & 6974 & 1.7x \\
uber      & 4 &  3,309,490 &  418 &  107 &   594 &  398 & 1.9x \\
vast-2015 & 5 & 26,021,945 & 3796 & 1728 &  5191 & 2832 & 1.5x \\
\hline
\end{tabular}
\lebat{Sparse Tensor Input (time in ms.)}{tab:sparseio}

\subsection{Sparse Linear Algebra}

To validate the quality of the generated sparse code, we measured the runtime of the SpMSpM kernel for \texttt{C[i,j] = A[i,k] * B[k,j]} that was discussed in Section~\ref{trafo}, using various $n \times n$ uniform random sparse matrices with a fixed density of $\rho_{A,B} = 0.01$, illustrated on the left in Figure~\ref{fig:inputs}, as input operands \texttt{A} and \texttt{B}. Table~\ref{tab:sparsematmul} summarizes the resulting density $\rho_C$ for output matrix \texttt{C}, the measured runtime of the kernel, as well as the relative performance improvement of MLIR over TACO. Since both sparse compilers generate more or less the same code, as expected, the performance is quite similar, with a minor speedup using MLIR, probably due to back-end optimization differences.

\plaat{bh}
\frame{\includegraphics[scale=0.2]{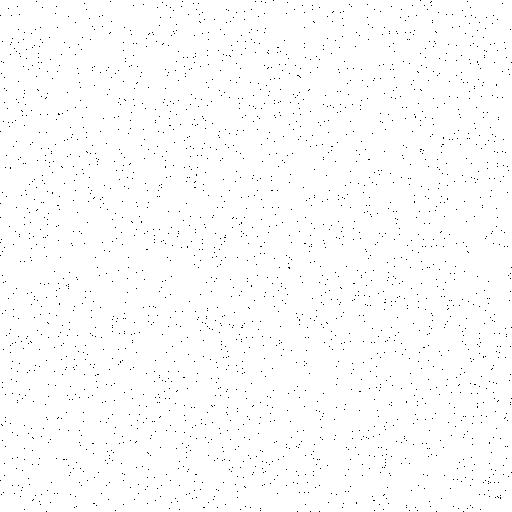}} \ \ 
\frame{\includegraphics[scale=0.2]{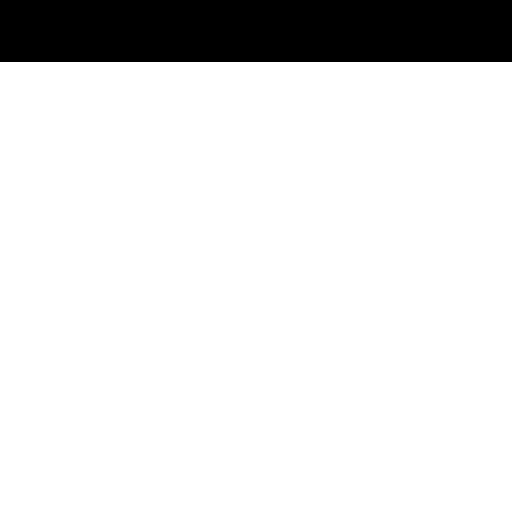}}
\taalp{Sparse Matrices (uniform random and row band)}{fig:inputs}

\tabel{bh}
\footnotesize
\begin{tabular}{|rr|r|r|l|}
\hline
$n$  & $\rho_C$ & \textbf{MLIR} & \textbf{TACO} & \\
\hline
1,024 & $0.10$ &     3.7 &     3.9 & 1.06x \\
2,048 & $0.19$ &    27.0 &    29.1 & 1.08x \\
4,096 & $0.34$ &   224.4 &   244.7 & 1.09x \\
8,192 & $0.56$ & 1,775.1 & 1,900.4 & 1.07x \\
\hline
\end{tabular}
\lebat{SpMSpM Kernel (time in ms.)}{tab:sparsematmul}

To demonstrate the impact of sparse storage format selection on performance, consider the following sparse matrix times vector operation, where we have a choice for the sparse type \texttt{SparseMatrix}.

\exam
\verb"%x = linalg.matvec ins(%A, %b: tensor<?x?xf64, #SparseMatrix>, tensor<?xf64>)-> tensor<?xf64>" \\
\maxe

Obvious choices for an unstructured sparse matrix \texttt{A} as input would be formats like CSR or DCSR, with the latter favoring sparse matrices that have many empty rows. However, for row band matrices, illustrated on the right in Figure~\ref{fig:inputs}, where many rows are empty, but non-empty rows are dense, using a type encoding with \texttt{compressed} for the outermost dimension and \texttt{dense} for the innermost dimension would make more sense. Lets call this format CDR. Table~\ref{tab:sparsematvec} compares the performance of these storage formats for the sparse code generated by MLIR and TACO for various $n \times n$ row band matrices where the first 1000 rows are kept dense, so that density $\rho_A$ decreases as size $n$ increases.

For MLIR, we also enabled the built-in vectorization strategy to take full advantage of AVX512 instructions on the target architecture along dense rows. As expected, CSR and DCSR perform similar for both compilers, with a slight advantage in the doubly compressed format due to skipping empty rows completely. For both compilers, however, the CDR version performs best. In addition, MLIR takes full advantage of the SIMD instructions, as indicated by the last column which shows the relative performance improvement of MLIR over TACO for the best version. This performance improvement demonstrates the utility of integrating sparse compilation into a modern multi-level compiler infrastructure.

\tabel{t}
\footnotesize
\begin{tabular}{|rr|rrr|rrr|l|}
\hline
\multicolumn{2}{|c|}{}              &
\multicolumn{3}{|c|}{\textbf{MLIR}} &
\multicolumn{3}{|c|}{\textbf{TACO}} &  \\
$n$ & $\rho_A$ & \textbf{CSR} & \textbf{DCSR} & \textbf{CDR} & \textbf{CSR} & \textbf{DCSR} & \textbf{CDR} & \\
\hline
 8,192 & 0.12 &  7.88 &  7.85 & 5.22 & 10.99 & 10.78  & 6.62  & 1.27x \\
16,384 & 0.06 & 15.76 & 15.69 & 10.52 & 19.77 & 19.21 & 11.90 & 1.13x \\
32,768 & 0.03 & 31.51 & 31.41 & 21.07 & 38.63 & 37.27 & 24.79 & 1.18x \\
65,536 & 0.02 & 77.28 & 72.19 & 42.18 & 77.55 & 76.84 & 47.07 & 1.12x \\
\hline
\end{tabular}
\lebat{SpMV Kernel (time in ms.)}{tab:sparsematvec}

Lastly, to illustrate the use of a sparse state space search to find a suitable sparse storage scheme, we conducted some experiments with the well-known sampled dense-dense matrix multiplication (SDDMM) kernel~\cite{sampleddd} that samples the result of a regular matrix multiplication by means of an element-wise multiplication with a sparse matrix, expressed in tensor index notation as \texttt{X[i,j] = S[i,j] * A[i,k] * B[k,j]} for a sparse matrix \texttt{S}. Figure~\ref{fig:space} plots the runtime of this kernel for different versions generated by MLIR using various sparse storage schemes for \texttt{S} (left-to-right) and compiler optimization strategies (sorted front-to-back). The figure clearly visualizes how an expert programmer can find the best storage format and optimization strategy by looking for the minimum runtime in the plot (here, DCSR and an optimization that corresponds to vectorization using 32-bit indices).

\plaat{ht}
\includegraphics[scale=0.3]{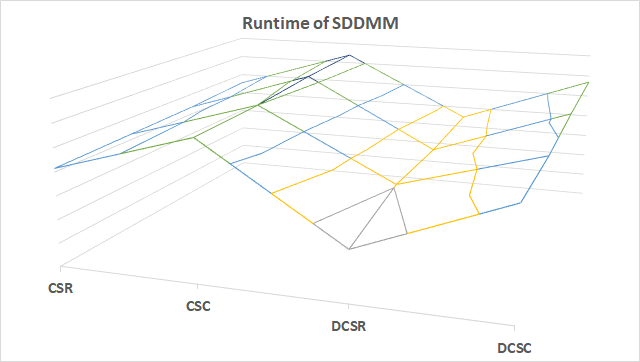}
\taalp{Sparse Runtime State Space for SDDMM (lower is better)}{fig:space}

\subsection{Sparse Tensor Algebra}

To validate the quality of the generated code for tensor algebra, we measured the runtime of Matricized Tensor Times Khatri-Rao Product, or MTTKRP for short. This kernel, expressed in tensor index notation as \texttt{A[i,j] = B[i,k,l] * D[l,j] * C[k,j]} (viz. two reduction loops) for a sparse 3-dimensional tensor $B$, forms the bottleneck of various problems in data analytics. Table~\ref{tab:sparsemttkrp} shows the runtime of this kernel for the $12092 \times 9184 \times 28818$ sparse tensor \texttt{nell-2} of FROSTT, used earlier in the input experiment, and size $12092 \times 42$ for the output matrix. For both MLIR and TACO, we measured the time for the sparse versions generated for formats with an outermost \texttt{dense}  and two innermost \texttt{compressed} dimensions and default lexicographic index order, as well as an alternative format with dimension ordering that interchanges the two innermost dimension (at the expense of a slightly higher packer time).
For MLIR, we also measured the runtime using a vectorization strategy for the latter format. The table clearly shows the advantage of the alternative dimension ordering, as well as the benefits obtained by using MLIR's optimization capabilities.

\tabel{t}
\footnotesize
\begin{tabular}{|r|rrr|rr|l|}
\hline              
& \multicolumn{3}{|c|}{\textbf{MLIR}} &
  \multicolumn{2}{|c|}{\textbf{TACO}} & \\
\textbf{tensor} & \textbf{DSS}  & \textbf{DSS-alt} &  \textbf{DSS-vec}  & \textbf{DSS} & \textbf{DSS-alt} & \\
\hline     
nell-2   & 2,340.82 & 1,754.98  & 1,299.97  & 2,587.81  & 1,871.64 & 1.44x \\
\hline
\end{tabular}
\lebat{MTTKRP Kernel (time in ms.)}{tab:sparsemttkrp}

\section{Related Work}
\label{related}

This section discusses related work, organized by sparsity support in compilers and libraries.

\subsection{Sparse Compilers}

To the best of our knowledge, Bik and Wijshoff were the first to propose the concept of treating sparsity merely as a property and letting a sparse compiler generate sparse code automatically from a sparsity-agnostic definition of the computation~\cite{bikjpdc}, which resulted in the \emph{MT1 sparse compiler} for linear algebra~\cite{drbik,biktpds}. This Fortran compiler first evaluates an attribute grammar to associate sparse conditions for execution with all expressions and statements in the input program. Then, driven by annotations and automatic nonzero structure analysis of representative sparse input matrices~\cite{biksicomp}, the program is automatically converted into a form that only stores and operates on nonzero elements for unstructured and structured sparse matrices. Using this sparse compiler for automatic sparse library generation was demonstrated in~\cite{biktms} for linear algebra primitives.

The concept of generating sparse code from dense abstractions is shared with subsequent work. \emph{The Vienna-Fortran / HPF sparse extensions}~\cite{ujaldon1997hpf-sparse-extension} provides High-Performance Fortran language features where users can characterize a matrix as sparse and specify the associated representation for its underlying compiler. \emph{The Bernoulli project}~\cite{kotlyar1997bernoulli,mateev2000generic-programming} generates efficient serial and distributed sparse matrix code from HPF-style, dense DO-ANY loops through relational algebra. By viewing arrays as relations and execution of loop nests as evaluation of relational queries, the problem of effectively traversing the sparse iteration space becomes finding optimal select and join schedules. Users have to manually describe desired sparse storage format by defining access methods for searching and enumerating indices. The \emph{SIPR framework}~\cite{pugh1998sipr} applies compilation to sparse algorithms with auxiliary data structures, such as a sparse accumulator~\cite{gilbert1992spa}. Leveraging the concepts of Aspect Oriented Programming, SIPR represents sparse programs as a combination of dense algorithms and dynamic sparse storage specifications, mapping array references and loop iterations to storage formats and array access descriptors. The latter specify an algorithm used to access the array elements, such as, direct access, enumeration, or binary search. The associated cost model allows for automatic computation of asymptotic complexity. The backend tool converts SIPR programs to C++ achieving performance on par with hand-written Fortran.

Sparse compilation was formalized and generalized to sparse tensor algebra in \emph{TACO (Tensor Algebra Compiler)}~\cite{taco,drkjolstad,taco2018,taco2017}. Starting with a ``loopless'' tensor index notation, the programmer annotates tensors as sparse. For this, TACO was the first to propose the elegant way of expressing desired sparse storage schemes with a combination of per-dimension level types and a dimension ordering. The dimension level types were expanded by Chou et al.~\cite{chou2018} to include more sparse storage formats. Kjolstad et al. were also the first to present a sparse iteration model that drives sparse code generation by topologically sorting the iteration graph to find a suitable index order, and using iteration lattices to emit loops, conditions, and tensor expressions for every index in topological order. For this, TACO  provides the co-iteration formulation that can be used to generate code to co-iterate over any number of sparse and dense tensors, which is necessary for general kernel fusion. Kjolstad et al.~\cite{taco2018} and Senanayake et al.~\cite{senanayake2020} also extended the TACO compiler with a scheduling language that lets users (or automatic systems) organize the iteration over tensor expressions, which lets them tile, control fusion/fission, statically load-balance, and generate GPU code for sparse tensor algebra kernels.  Sparse tensor support in MLIR borrows heavily from the foundation laid by TACO.

\emph{Taichi}~\cite{hu2019taichi} is a programming language that offers a data structure-agnostic interface for writing computation code. The user specifies the data structure with different sparsity properties independent of the code to create a wide range of sparse data structures. As with the approaches above, decoupling data structures from computation simplifies experimenting with different data structures without the need to change the actual computation code.

\emph{COMET}~\cite{mutlu2021comet, tian2021comet} is a MLIR-based compiler infrastructure for dense and sparse tensor algebra computations. It uses a dimension-wise sparse storage scheme and a code generation algorithm similar to TACO, but has more performance portability building upon MLIR. Its DSL can succinctly describe tensor algebra expressions using Einstein summation notation and format tags for sparse tensors. The DSL is mapped to a custom Sparse Tensor Algebra dialect, which is progressively lowered to LLVM IR through MLIR dialects. The framework also incorporates a data reordering algorithm to help with data locality. Among compiler-based approaches, COMET's is the closest to ours.

\emph{Henry et al.}~\cite{henry2021compilation} generalized the sparse tensor algebra compilation theory from TACO~\cite{taco2017} to support compilation of general dense and sparse array expressions, laying the foundation for a sparse NumPy-style system. In this system, arrays (tensors) can have any implicit fill value and any function can be computed across sparse and dense arrays. The new scheme covers sparse iteration spaces outside of union (addition) and intersection (multiplication) for arbitrary user-defined functions. It can also iterate over slices or strides of sparse arrays, allowing for a much wider range of kernels and applications.

Another line of work is to directly optimize sparse code, i.e., codes that deal with compressed data structures. \emph{SPARSITY}~\cite{im2004sparsity} provides a framework that selects optimization parameters to automatically build sparse matrix kernels that are tuned to their matrices and machines. The framework combines traditional loop transformations with data structure transformations that are specific to sparse matrices. \emph{LL (Little Language)}~\cite{arnold2011ll} is a small functional language aiming to increase programming productivity of sparse operations. It provides built-in nested list and pair types, with which users can naturally represent compressed matrices in many formats. Sparse computations are defined through a data flow graph. The compiler can generate efficient, fully verified C code.

\emph{The Sparse Polyhedral Framework (SPF)}~\cite{strout2018spf} provides code and data transformations for non-affine loop bounds and array index expressions~\cite{venkat2015loop} which can be combined with existing transformation in polyhedral framework. Each transformation controls whether a dimension in the sparse iteration space is traversed fully (\emph{make-dense}) or sparsely (\emph{compact}), with an option to pad the underlying data storage for that dimension (\emph{compact-and-pad}). The generated code is akin to converting sparse tensor to a specific layout then performing the computation. SPF can cover a large variety of tensor storage formats, at the expense of users having to write a sequence of transformations to produce the desired computations.

\emph{TIRAMISU}~\cite{baghdadi2020tiramisu} is a polyhedral compiler for deep learning that maps sparse and recurrent neural networks to the polyhedral model in order to generate highly-optimized sparse code for the target architecture.

\subsection{Sparse Libraries and Kernels}

A large number of applications build upon sparse computation library primitives optimized for their target architectures. There are numerous sparse libraries for different application domains with varying performance and productivity goals.

Classic sparse linear algebra binary libraries like \emph{MKL}~\cite{wang2014mkl} and \emph{cuSPARSE}~\cite{naumov2010cusparse} implements sparse basic linear algebra subroutines with a few data types. More recent generic C++ libraries such as \emph{Eigen}~\cite{guennebaud2010eigen} and \emph{CUSP}~\cite{dalton2014cusp} allow writing math-like expressions and can support more data types. \emph{OSKI}~\cite{vuduc2005oski} library offers automatically-tuned sparse matrix CPU primitives for linear solvers, with an easy-to-use interface.

\emph{PETSc}~\cite{balay1997petsc} is a library of scalable PDE solvers that has modular structure and multiple abstraction layers, making it easy to add support for new devices, kernels, data types, etc., including new sparse matrix formats~\cite{kumbhar2011petsc-sparse-format-extension}. It can also be integrated with a compiler-based autotuning approach to further optimize its hotspot kernels~\cite{ramalingam2012petsc-compiler}.

Matrix algebra can also be beneficial in graph processing~\cite{harary1969graph}.  The \emph{GraphBLAS}~\cite{kepner2015graphblas-standard} standard specifies a core set of general sparse matrix-based graph operations, e.g., generalized matrix operations over arbitrary semirings (replacing element-wise addition and multiplication with other operators), providing additional operation coverage to common sparse linear algebra libraries. Libraries implementing this standard includes Combinatorial BLAS~\cite{buluc2011cblas}, GraphPad~\cite{anderson2016graphpad}, GraphBLAST~\cite{yang2019graphblast}, SuiteSparse:GraphBLAS~\cite{davis2019suitesparse-graphblas}, etc.

Outside of deep learning, most common sparse tensor computations are tensor decompositions~\cite{kolda2009tensor-decomposition} and contractions~\cite{parkhill2010sparse-contraction, kats2013sparse-contraction}. The \emph{Cyclops Tensor Framework}~\cite{solomonik2013ctf, solomonik2015sparse-ctf} is a C++ template library for distributed dense and sparse tensor algebra targeting quantum chemistry domain. Through operator overloading, users can write tensor algebra expression with Einstein summation notation directly in C++, e.g., \texttt{C["hij"] += A["ijhk"] * B["hkij"]}. It supports templated data types and can substitute element-wise operations in tensor addition and contraction with other operations through user-defined algebraic structures (e.g., semirings, groups, etc). CTF's execution strategy is to transform tensors into a layout that fits hand-implemented library routines, including a 2.5D matrix multiplication, invoke those routines, and then transform the data back. Among library-based approaches, CTF is the closest related work to ours due to its generality.

The rising popularity of deep learning poses new challenges to sparse libraries, e.g., new sparsity characteristics, kernels, data types, hardware accelerators, etc. While there are libraries such as Sputnik~\cite{gale2020sputnik}, cuSPARSELt~\cite{nvidia2021cusparselt}, and LIBXSMM~\cite{heinecke2016libxsmm} adding new kernels and data types specific to deep learning, these kernels still have limited composability and portability. 

Specializing sparse kernels for specific sparsity characteristics or hardware architecture can bring significant performance gains. Numerous applications implement custom sparse formats that better suit their sparsity patterns and target hardware~\cite{sato1963,tinney,buluc2008,itpack,rice1985,saad2003,buluc2009,smith2015a,li2018hicoo}. Recent studies~\cite{zhao2018spmv-format,pichel2019classification,xie2019ia-spgemm} on sparse matrix format classification uses machine learning determine the best input storage format to maximize the performance of a particular kernel. Only a small number of matrix formats were compared, because each of them needs separate implementation. With our work, an arbitrarily large number of formats can simply be explored as compiler tuning knobs~\cite{gus}.

\section{Conclusions and Future Plans}
\label{conclusions}

In this paper, we proposed treating sparsity as a property, not a tedious implementation task, and showed how the concept of letting a sparse compiler generate sparse code automatically from a sparsity-agnostic definition of the computation is being integrated into the MLIR open-source compiler infrastructure. MLIR's ease of defining multi-level abstractions and transformations composes well with this idea of making sparsity a property. We discussed our north star vision of providing an excellent, reusable sparse ecosystem by introducing sparse tensor types as proper first-class citizens into MLIR. A reference implementation provides a fully functional implementation of this sparse ecosystem. We discussed two ways of using sparse compiler support in MLIR, namely, end-to-end support for array languages and state space search for sparse library development. Lastly, we provided some experimental validation of our initial reference implementation.

Going forward, our hope is that the open-source nature of this project will not only benefit researchers that are new to the sparse domain but also solicit useful contributions from experts in the open-source community at large. Many interesting topics remain under active research and development. For instance, our current sparsity encoding with just dense and compressed dimension level types can be generalized to allow other dimension level types as well, which will widen the scope of the current implementation from unstructured and block-structured sparse tensors into new unstructured and structured sparse tensor storage formats that better exploit specific features of the tensors, the computation, and the target architecture. Also, mapping general array language constructs into a suitable intermediate representation is under active investigation. Furthermore, even though MLIR has the ability to apply fusion, fission, tiling, and various other optimizations to \texttt{Linalg} operations, more work is needed on compiler strategies for doing so in the context of sparse tensors. More research is also needed to generate efficient code for kernels with complex access patterns, such as sparse convolutions. The introduction of set-oriented looping constructs to MLIR would enable a more gradual progressive lowering into sparse code, since the current direct lowering to \texttt{SCF} bridges a rather wide semantic gap. Finally, we plan to extend the scope of our reference implementation from CPUs to GPUs, TPUs, and other kinds of accelerators eventually.

The MLIR project was in part started to provide an extensible and powerful compiler infrastructure for domain-specific languages. Of particular interest was dense tensor algebra for deep neural networks. But while dense tensor computations remain the bulk of machine learning workloads, sparse neural networks are increasingly being explored, whether weight-sparse, activation-sparse, or input sparse such as graph-convolutional neural networks.

With the MLIR sparse compiler we seek to put compilation for sparse tensor algebra on the same strong footing as compilation for dense tensor algebra. We believe this is necessary to enable both efficient exploration and efficient computation of sparse computations in machine learning and in data analytics. In fact, classical compiler transformations for dense tensor computations, such as reordering, fission, and fusion, can be even more important for sparse computation, as getting loop ordering and fusion/fission wrong often leads to \emph{asymptotically worse complexity}. This effect occurs even in staple computations like sparse matrix-matrix multiplication and sampled dense-dense matrix multiplication, which are used in sparse neural networks. The MLIR sparse compiler is positioned to address these challenges and we are therefore excited about its potential.

\textbf{Acknowledgments} The authors would like to extend their thanks to the MLIR team, with a special shout-out to Mehdi Amini, Eugene Burmako, Diego Caballero, Albert Cohen, Sanjoy Das, Tobias Gysi, Stephan Herhut, Stella Laurenzo, Jacques Pienaar, Thomas Raoux, River Riddle, Wren Romano, Sean Silva, Gus Smith, Matthias Springer, Reid Tatge, Jake van der Plas, Alex Zinenko, and Eugene Zhulenev.

\bibliographystyle{acm}
\bibliography{mlir}

\begin{thebibliography}{10}

\bibitem{anderson}
{\sc Anderson, E., and Saad, Y.}
\newblock Solving sparse triangular linear systems on parallel computers.
\newblock {\em International Journal of High Speed Computing 1}, 6 (1989),
  73--95.

\bibitem{anderson2016graphpad}
{\sc Anderson, M.~J., Sundaram, N., Satish, N., Patwary, M. M.~A., Willke,
  T.~L., and Dubey, P.}
\newblock Graphpad: Optimized graph primitives for parallel and distributed
  platforms.
\newblock In {\em 2016 IEEE International Parallel and Distributed Processing
  Symposium (IPDPS)\/} (2016), IEEE, pp.~313--322.

\bibitem{arnold2011ll}
{\sc Arnold, G.}
\newblock {\em Data-Parallel Language for Correct and Efficient Sparse Matrix
  Codes}.
\newblock University of California, Berkeley, 2011.

\bibitem{baghdadi2020tiramisu}
{\sc Baghdadi, R., Debbagh, A.~N., Abdous, K., Benhamida, F.~Z., Renda, A.,
  Frankle, J.~E., Carbin, M., and Amarasinghe, S.}
\newblock Tiramisu: A polyhedral compiler for dense and sparse deep learning.
\newblock {\em arXiv preprint arXiv:2005.04091\/} (2020).

\bibitem{balay1997petsc}
{\sc Balay, S., Gropp, W.~D., McInnes, L.~C., and Smith, B.~F.}
\newblock Efficient management of parallelism in object-oriented numerical
  software libraries.
\newblock In {\em Modern software tools for scientific computing}. Springer,
  1997, pp.~163--202.

\bibitem{drbik}
{\sc Bik, A.~J.}
\newblock {\em Compiler Support for Sparse Matrix Computations}.
\newblock PhD thesis, Department of Computer Science, Leiden University, 1996.
\newblock ISBN 90-9009442-3.

\bibitem{biktms}
{\sc Bik, A.~J., Brinkhaus, P.~J., Knijnenburg, P.~M., and Wijshoff, H.~A.}
\newblock The automatic generation of sparse primitives.
\newblock {\em Transactions on Mathematical Software 24\/} (1998), 190--225.

\bibitem{bikjpdc}
{\sc Bik, A.~J., and Wijshoff, H.~A.}
\newblock Advanced compiler optimizations for sparse computations.
\newblock {\em Journal of Parallel and Distributed Computing 31\/} (1995),
  14--24.

\bibitem{biktpds}
{\sc Bik, A.~J., and Wijshoff, H.~A.}
\newblock Automatic data structure selection and transformation for sparse
  matrix computations.
\newblock {\em IEEE Transactions on Parallel and Distributed Systems 7}, 2
  (1996), 109--126.

\bibitem{biksicomp}
{\sc Bik, A.~J., and Wijshoff, H.~A.}
\newblock Automatic nonzero structure analysis.
\newblock {\em SIAM J. Computing 28}, 5 (1999), 1576--1587.

\bibitem{matrixmarket}
{\sc Boisvert, R., Pozo, R., and Remington, K.}
\newblock The {M}atrix {M}arket {E}xchange {F}ormats: Initial design, 1996.
\newblock NIST Interagency/Internal Report (NISTIR), National Institute of
  Standards and Technology, Gaithersburg, MD.

\bibitem{buluc2009}
{\sc Bulu{\c{c}}, A., Fineman, J.~T., Frigo, M., Gilbert, J.~R., and Leiserson,
  C.~E.}
\newblock Parallel sparse matrix-vector and matrix-transpose-vector
  multiplication using compressed sparse blocks.
\newblock In {\em ACM Symposium on Parallelism in Algorithms and
  Architectures\/} (New York, NY, USA, 2009), ACM, p.~233.

\bibitem{buluc2008}
{\sc Bulu{\c{c}}, A., and Gilbert, J.~R.}
\newblock On the representation and multiplication of hypersparse matrices.
\newblock In {\em {IEEE} International Symposium on Parallel and Distributed
  Processing, (IPDPS).\/} (Apr. 2008), pp.~1--11.

\bibitem{buluc2011cblas}
{\sc Bulu{\c{c}}, A., and Gilbert, J.~R.}
\newblock The combinatorial blas: Design, implementation, and applications.
\newblock {\em The International Journal of High Performance Computing
  Applications 25}, 4 (2011), 496--509.

\bibitem{chou2018}
{\sc Chou, S., Kjolstad, F., and Amarasinghe, S.}
\newblock Format abstraction for sparse tensor algebra compilers.
\newblock {\em Proc. ACM Program. Lang. 2}, OOPSLA (Oct. 2018), 123:1--123:30.

\bibitem{chou2020}
{\sc Chou, S., Kjolstad, F., and Amarasinghe, S.}
\newblock Automatic generation of efficient sparse tensor format conversion
  routines.
\newblock In {\em Proceedings of the 41st ACM SIGPLAN Conference on Programming
  Language Design and Implementation\/} (New York, NY, USA, 2020), PLDI 2020,
  Association for Computing Machinery, pp.~823--838.

\bibitem{coleman}
{\sc Coleman, T.~F.}
\newblock Large sparse numerical optimization.
\newblock In {\em Lecture Notes in Computer Science, No.~165}, G.~Goos and
  J.~Hartmanis, Eds. Springer-Verlag, Berlin, 1984.

\bibitem{nvidia2021cusparselt}
{\sc Corporation, N.}
\newblock cusparselt: A high-performance cuda library for sparse matrix-matrix
  multiplication, 2021.

\bibitem{curtis71}
{\sc Curtis, A., and Reid, J.}
\newblock The solution of large sparse unsymmetric systems of linear equations.
\newblock {\em Journal Inst. Maths. Applics. 8\/} (1971), 344--353.

\bibitem{dalton2014cusp}
{\sc Dalton, S., Bell, N., Olson, L., and Garland, M.}
\newblock Cusp: Generic parallel algorithms for sparse matrix and graph
  computations, 2014.
\newblock Version 0.5.0.

\bibitem{davis2019suitesparse-graphblas}
{\sc Davis, T.~A.}
\newblock Algorithm 1000: Suitesparse: Graphblas: Graph algorithms in the
  language of sparse linear algebra.
\newblock {\em ACM Transactions on Mathematical Software (TOMS) 45}, 4 (2019),
  1--25.

\bibitem{duffo}
{\sc Duff, I.~S.}
\newblock A survey of sparse matrix research.
\newblock In {\em Proceedings of the IEEE\/} (1977), pp.~500--535.

\bibitem{duff5}
{\sc Duff, I.~S.}
\newblock Data structures, algorithms and software for sparse matrices.
\newblock In {\em Sparsity and Its Applications}, D.~J. Evans, Ed. Cambridge
  University Press, 1985, pp.~1--29.

\bibitem{duff}
{\sc Duff, I.~S., Erisman, A., and Reid, J.}
\newblock {\em Direct Methods for Sparse Matrices}.
\newblock Oxford Science Publications, Oxford, 1990.

\bibitem{duff2}
{\sc Duff, I.~S., Grimes, R.~G., and Lewis, J.~G.}
\newblock Sparse matrix test problems.
\newblock {\em ACM Transactions on Mathematical Software 15\/} (1989), 1--14.

\bibitem{evans}
{\sc Evans, D.}
\newblock Iterative sparse matrix algorithms.
\newblock In {\em Software for Numerical Mathematics}, D.~Evans, Ed. Academic
  Press, New York, NY, 1974, pp.~49--83.

\bibitem{gale2020sputnik}
{\sc Gale, T., Zaharia, M., Young, C., and Elsen, E.}
\newblock Sparse {GPU} kernels for deep learning.
\newblock In {\em Proceedings of the International Conference for High
  Performance Computing, Networking, Storage and Analysis, {SC} 2020\/} (2020).

\bibitem{george}
{\sc George, A., and Liu, J.~W.}
\newblock {\em Computer Solution of Large Sparse Positive Definite Systems}.
\newblock Prentice Hall, Englewood Cliffs, New York, 1981.

\bibitem{gilbert1992spa}
{\sc Gilbert, J., Moler, C., and Schreiber, R.}
\newblock Sparse matrices in matlab: Design and implementation.
\newblock {\em SIAM J. on Matrix Analysis and Applications 13}, 1 (1992),
  333--356.

\bibitem{guennebaud2010eigen}
{\sc Guennebaud, G., Jacob, B., et~al.}
\newblock Eigen.
\newblock {\em URl: http://eigen. tuxfamily. org 3\/} (2010).

\bibitem{rose}
{\sc Gustavson, F.~G.}
\newblock Some basic techniques for solving sparse systems of linear equations.
\newblock In {\em Sparse Matrices and Their Applications}, D.~J. Rose and R.~A.
  Willoughby, Eds. Plenum Press, New York, NY, 1972, pp.~41--52.

\bibitem{harary1969graph}
{\sc Harary, F.}
\newblock Graph theory, chs. 2, 13.
\newblock {\em Reading: Addison Wesley\/} (1969).

\bibitem{harary}
{\sc Harary, F.}
\newblock Sparse matrices and graph theory.
\newblock In {\em Large Sparse Sets of Linear Equations}, J.~Reid, Ed. Academic
  Press, 1971, pp.~139--150.

\bibitem{heinecke2016libxsmm}
{\sc Heinecke, A., Henry, G., Hutchinson, M., and Pabst, H.}
\newblock Libxsmm: accelerating small matrix multiplications by runtime code
  generation.
\newblock In {\em SC'16: Proceedings of the International Conference for High
  Performance Computing, Networking, Storage and Analysis\/} (2016), IEEE,
  pp.~981--991.

\bibitem{henry2021compilation}
{\sc Henry, R., Hsu, O., Yadav, R., Chou, S., Olukotun, K., Amarasinghe, S.,
  and Kjolstad, F.}
\newblock Compilation of sparse array programming models.
\newblock {\em Proceedings of the ACM on Programming Languages 5}, OOPSLA
  (2021), 1--29.

\bibitem{hu2019taichi}
{\sc Hu, Y., Li, T.-M., Anderson, L., Ragan-Kelley, J., and Durand, F.}
\newblock Taichi: a language for high-performance computation on spatially
  sparse data structures.
\newblock {\em ACM Transactions on Graphics (TOG) 38}, 6 (2019), 1--16.

\bibitem{im2004sparsity}
{\sc Im, E.-J., Yelick, K., and Vuduc, R.}
\newblock Sparsity: Optimization framework for sparse matrix kernels.
\newblock {\em The International Journal of High Performance Computing
  Applications 18}, 1 (2004), 135--158.

\bibitem{kats2013sparse-contraction}
{\sc Kats, D., and Manby, F.~R.}
\newblock Sparse tensor framework for implementation of general local
  correlation methods.
\newblock {\em The Journal of Chemical Physics 138}, 14 (2013), 144101.

\bibitem{kepner2015graphblas-standard}
{\sc Kepner, J., Bader, D., Bulu{\c{c}}, A., Gilbert, J., Mattson, T., and
  Meyerhenke, H.}
\newblock Graphs, matrices, and the graphblas: Seven good reasons.
\newblock {\em Procedia Computer Science 51\/} (2015), 2453--2462.

\bibitem{drkjolstad}
{\sc Kjolstad, F.}
\newblock {\em Sparse Tensor Algebra Compilation}.
\newblock PhD thesis, Massachusetts Institute of Technology, Cambridge, MA, Feb
  2020.

\bibitem{taco2018}
{\sc Kjolstad, F., Ahrens, P., Kamil, S., and Amarasinghe, S.}
\newblock Tensor algebra compilation with workspaces.
\newblock {\em Proceedings of the 2019 IEEE/ACM International Symposium on Code
  Generation and Optimization\/} (2019), 180--192.

\bibitem{taco}
{\sc Kjolstad, F., et~al.}
\newblock {TACO}: The tensor algebra compiler, 2017.
\newblock Open-source project available at \url{http://tensor-compiler.org/}.

\bibitem{taco2017}
{\sc Kjolstad, F., Kamil, S., Chou, S., Lugato, D., and Amarasinghe, S.}
\newblock The tensor algebra compiler.
\newblock {\em Proc. ACM Program. Lang. 1}, OOPSLA (Oct. 2017), 77:1--77:29.

\bibitem{kolda2009tensor-decomposition}
{\sc Kolda, T.~G., and Bader, B.~W.}
\newblock Tensor decompositions and applications.
\newblock {\em SIAM review 51}, 3 (2009), 455--500.

\bibitem{kotlyar1997bernoulli}
{\sc Kotlyar, V., Pingali, K., and Stodghill, P.}
\newblock A relational approach to the compilation of sparse matrix programs.
\newblock In {\em European Conference on Parallel Processing\/} (1997),
  Springer, pp.~318--327.

\bibitem{kumbhar2011petsc-sparse-format-extension}
{\sc Kumbhar, P.}
\newblock {\em Performance of petsc gpu implementation with sparse matrix
  storage schemes}.
\newblock PhD thesis, Master’s thesis, The University of Edinburgh (Aug
  2011), 2011.

\bibitem{mlir2021}
{\sc Lattner, C., Amini, M., Bondhugula, U., Cohen, A., Davis, A., Pienaar, J.,
  Riddle, R., Shpeisman, T., Vasilache, N., and Zinenko, O.}
\newblock {MLIR}: Scaling compiler infrastructure for domain specific
  computation.
\newblock In {\em 2021 IEEE/ACM International Symposium on Code Generation and
  Optimization (CGO)\/} (2021), pp.~2--14.

\bibitem{mlir2020}
{\sc Lattner, C., Pienaar, J.~A., Amini, M., Bondhugula, U., Riddle, R., Cohen,
  A., Shpeisman, T., Davis, A., Vasilache, N., and Zinenko, O.}
\newblock {MLIR:} {A} compiler infrastructure for the end of moore's law.
\newblock {\em CoRR abs/2002.11054\/} (2020).

\bibitem{li2018hicoo}
{\sc Li, J., Sun, J., and Vuduc, R.}
\newblock Hicoo: Hierarchical storage of sparse tensors.
\newblock In {\em SC18: International Conference for High Performance
  Computing, Networking, Storage and Analysis\/} (2018), IEEE, pp.~238--252.

\bibitem{liu}
{\sc Liu, J.~W.}
\newblock A compact row storage scheme for cholesky factors using elimination
  trees.
\newblock {\em ACM Transactions on Mathematical Software\/} (1986), 127--148.

\bibitem{mann}
{\sc Mann, K.~J.}
\newblock Inversion of large sparse matrices: Direct methods.
\newblock In {\em Numerical Solutions of Partial Differential Equations},
  J.~Noye, Ed. North-Holland Publishing Company, Amsterdam, 1982, pp.~313--366.

\bibitem{mateev2000generic-programming}
{\sc Mateev, N., Pingali, K., Stodghill, P., and Kotlyar, V.}
\newblock Next-generation generic programming and its application to sparse
  matrix computations.
\newblock In {\em Proceedings of the 14th international conference on
  Supercomputing\/} (2000), pp.~88--99.

\bibitem{mutlu2021comet}
{\sc Mutlu, E., Tian, R., Ren, B., Krishnamoorthy, S., Gioiosa, R., Pienaar,
  J., and Kestor, G.}
\newblock {COMET}: A domain-specific compilation of high-performance
  computational chemistry.
\newblock {\em arXiv preprint arXiv:2102.06827\/} (2021).

\bibitem{naumov2010cusparse}
{\sc Naumov, M., Chien, L., Vandermersch, P., and Kapasi, U.}
\newblock Cusparse library.
\newblock In {\em GPU Technology Conference\/} (2010).

\bibitem{sampleddd}
{\sc Nisa, I., Sukumaran-Rajam, A., Kurt, S.~E., Hong, C., and Sadayappan, P.}
\newblock Sampled dense matrix multiplication for high-performance machine
  learning.
\newblock In {\em 2018 IEEE 25th International Conference on High Performance
  Computing (HiPC)\/} (2018), pp.~32--41.

\bibitem{itpack}
{\sc Oppe, T.~C., and Kincaid, D.~R.}
\newblock The performance of {ITPACK} on vector computers for solving large
  sparse linear systems arising in sample oil reservoir simulation problems.
\newblock {\em Communications in Applied Numerical Methods 3}, 1 (1987),
  23--29.

\bibitem{parkhill2010sparse-contraction}
{\sc Parkhill, J.~A., and Head-Gordon, M.}
\newblock A sparse framework for the derivation and implementation of fermion
  algebra.
\newblock {\em Molecular Physics 108}, 3-4 (2010), 513--522.

\bibitem{pichel2019classification}
{\sc Pichel, J.~C., and Pateiro-Lopez, B.}
\newblock Sparse matrix classification on imbalanced datasets using
  convolutional neural networks.
\newblock {\em IEEE Access 7\/} (2019), 82377--82389.

\bibitem{sergio}
{\sc Pissanetsky, S.}
\newblock {\em Sparse Matrix Technology}.
\newblock Academic Press, London, 1984.

\bibitem{pugh1998sipr}
{\sc Pugh, W., and Shpeisman, T.}
\newblock Sipr: A new framework for generating efficient code for sparse matrix
  computations.
\newblock In {\em International Workshop on Languages and Compilers for
  Parallel Computing\/} (1998), Springer, pp.~213--229.

\bibitem{ramalingam2012petsc-compiler}
{\sc Ramalingam, S., Hall, M., and Chen, C.}
\newblock Improving high-performance sparse libraries using compiler-assisted
  specialization: A petsc case study.
\newblock In {\em 2012 IEEE 26th International Parallel and Distributed
  Processing Symposium Workshops \& PhD Forum\/} (2012), IEEE, pp.~487--496.

\bibitem{reid2}
{\sc Reid, J.}
\newblock Direct methods for sparse matrices.
\newblock In {\em Software for Numerical Mathematics}, D.~Evans, Ed. Academic
  Press, New York, NY, 1974, pp.~29--47.

\bibitem{rice1985}
{\sc Rice, J.~R., and Boisvert, R.~F.}
\newblock {\em Solving Elliptic Problems Using {ELLPACK}}.
\newblock Springer-Verlag, 1985.

\bibitem{saad2}
{\sc Saad, Y.}
\newblock {\em {SPARSKIT}: a basic tool kit for sparse matrix computations},
  1990.
\newblock CSRD/RIACS.

\bibitem{saad2003}
{\sc Saad, Y.}
\newblock {\em Iterative Methods for Sparse Linear Systems}.
\newblock SIAM, 2003.

\bibitem{sato1963}
{\sc Sato, N., and Tinney, W.~F.}
\newblock Techniques for exploiting the sparsity of the network admittance
  matrix.
\newblock {\em {IEEE} Transactions on Power Apparatus and Systems 82}, 69
  (1963), 944--950.

\bibitem{senanayake2020}
{\sc Senanayake, R., Hong, C., Wang, Z., Wilson, A., Chou, S., Kamil, S.,
  Amarasinghe, S., and Kjolstad, F.}
\newblock A sparse iteration space transformation framework for sparse tensor
  algebra.
\newblock {\em Proc. ACM Program. Lang. 4}, OOPSLA (Nov. 2020).

\bibitem{gus}
{\sc Smith, G.~H., Bik, A.~J., Koanantakool, P., and Phothilimthana, P.~M.}
\newblock {ML}-driven auto-configurator for sparse tensor kernels in mlir,
  2022.
\newblock Unpublished Manuscript.

\bibitem{frostt}
{\sc Smith, S., Choi, J.~W., Li, J., Vuduc, R., Park, J., Liu, X., and Karypis,
  G.}
\newblock {FROSTT}: The formidable repository of open sparse tensors and tools,
  2017.
\newblock \url{http://frostt.io/}.

\bibitem{fiber}
{\sc Smith, S., and Karypis, G.}
\newblock Tensor-matrix products with a compressed sparse tensor.
\newblock In {\em Proceedings of the 5th Workshop on Irregular Applications:
  Architectures and Algorithms\/} (New York, NY, USA, 2015), IA${}^3$ '15,
  Association for Computing Machinery.

\bibitem{smith2015a}
{\sc Smith, S., and Karypis, G.}
\newblock Tensor-matrix products with a compressed sparse tensor.
\newblock In {\em Workshop on Irregular Applications: Architectures and
  Algorithms\/} (2015), ACM, pp.~1--7.

\bibitem{solomonik2015sparse-ctf}
{\sc Solomonik, E., and Hoefler, T.}
\newblock Sparse tensor algebra as a parallel programming model.
\newblock {\em arXiv preprint arXiv:1512.00066\/} (2015).

\bibitem{solomonik2013ctf}
{\sc Solomonik, E., Matthews, D., Hammond, J., and Demmel, J.}
\newblock Cyclops tensor framework: Reducing communication and eliminating load
  imbalance in massively parallel contractions.
\newblock In {\em 2013 IEEE 27th International Symposium on Parallel and
  Distributed Processing\/} (2013), IEEE, pp.~813--824.

\bibitem{strout2018spf}
{\sc Strout, M.~M., Hall, M., and Olschanowsky, C.}
\newblock The sparse polyhedral framework: Composing compiler-generated
  inspector-executor code.
\newblock {\em Proceedings of the IEEE 106}, 11 (2018), 1921--1934.

\bibitem{tew2016}
{\sc Tew, P.~A.}
\newblock An investigation of sparse tensor formats for tensor libraries.
\newblock M.eng. thesis, Massachusetts Institute of Technology, Cambridge, MA,
  Jun 2016.

\bibitem{tewarson}
{\sc Tewarson, R.~P.}
\newblock {\em Sparse Matrices}.
\newblock Academic Press, New York, NY, 1973.

\bibitem{tian2021comet}
{\sc Tian, R., Guo, L., Li, J., Ren, B., and Kestor, G.}
\newblock A high performance sparse tensor algebra compiler in {MLIR}.
\newblock In {\em 2021 IEEE/ACM 7th Workshop on the LLVM Compiler
  Infrastructure in HPC (LLVM-HPC)\/} (2021), pp.~27--38.

\bibitem{tinney}
{\sc Tinney, W.~F., and Walker, J.~W.}
\newblock Direct solutions of sparse network equations by optimally ordered
  triangular factorization.
\newblock In {\em Proceedings of the IEEE\/} (1967), pp.~1801--1809.

\bibitem{ujaldon1997hpf-sparse-extension}
{\sc Ujaldon, M., Zapata, E.~L., Chapman, B.~M., and Zima, H.~P.}
\newblock Vienna-fortran/hpf extensions for sparse and irregular problems and
  their compilation.
\newblock {\em IEEE Transactions on Parallel and Distributed Systems 8}, 10
  (1997), 1068--1083.

\bibitem{nicolas2018}
{\sc Vasilache, N., Zinenko, O., Theodoridis, T., Goyal, P., DeVito, Z., Moses,
  W.~S., Verdoolaege, S., Adams, A., and Cohen, A.}
\newblock Tensor comprehensions: Framework-agnostic high-performance machine
  learning abstractions.
\newblock {\em CoRR abs/1802.04730\/} (2018).

\bibitem{marinus}
{\sc Veldhorst, M.}
\newblock {\em An Analysis of Sparse Matrix Storage Schemes}.
\newblock PhD thesis, Mathematisch Centrum, Amsterdam, 1982.

\bibitem{venkat2015loop}
{\sc Venkat, A., Hall, M., and Strout, M.}
\newblock Loop and data transformations for sparse matrix code.
\newblock {\em ACM SIGPLAN Notices 50}, 6 (2015), 521--532.

\bibitem{vuduc2005oski}
{\sc Vuduc, R., Demmel, J.~W., and Yelick, K.~A.}
\newblock {OSKI}: A library of automatically tuned sparse matrix kernels.
\newblock {\em Journal of Physics: Conference Series 16\/} (jan 2005),
  521--530.

\bibitem{wang2014mkl}
{\sc Wang, E., Zhang, Q., Shen, B., Zhang, G., Lu, X., Wu, Q., and Wang, Y.}
\newblock Intel math kernel library.
\newblock In {\em High-Performance Computing on the Intel{\textregistered} Xeon
  Phi™}. Springer, 2014, pp.~167--188.

\bibitem{xie2019ia-spgemm}
{\sc Xie, Z., Tan, G., Liu, W., and Sun, N.}
\newblock Ia-spgemm: An input-aware auto-tuning framework for parallel sparse
  matrix-matrix multiplication.
\newblock In {\em Proceedings of the ACM International Conference on
  Supercomputing\/} (2019), pp.~94--105.

\bibitem{yang2019graphblast}
{\sc Yang, C., Buluc, A., and Owens, J.~D.}
\newblock Graphblast: A high-performance linear algebra-based graph framework
  on the gpu.
\newblock {\em arXiv preprint arXiv:1908.01407\/} (2019).

\bibitem{zhao2018spmv-format}
{\sc Zhao, Y., Li, J., Liao, C., and Shen, X.}
\newblock Bridging the gap between deep learning and sparse matrix format
  selection.
\newblock In {\em Proceedings of the 23rd ACM SIGPLAN symposium on principles
  and practice of parallel programming\/} (2018), pp.~94--108.

\bibitem{zlatev}
{\sc Zlatev, Z.}
\newblock {\em Computational Methods for General Sparse Matrices}.
\newblock Kluwer Academic Publishers, Dordrecht, 1991.

\end{thebibliography}

\end{document}